\documentclass[letterpaper,twocolumn,10pt]{article}
\PassOptionsToPackage{hyphens}{url}
\usepackage{usenix2019_v3}
\usepackage{cite}
\usepackage[hyphens]{url}
\usepackage{balance}
\usepackage{booktabs}
\usepackage{tabularx}
\usepackage{makecell}
\usepackage{multirow}
\usepackage{diagbox}
\usepackage{subcaption}
\usepackage{threeparttable}
\usepackage{algpseudocode}
\usepackage{amsmath}
\usepackage[linesnumbered,ruled,vlined]{algorithm2e}
\usepackage{color}
\usepackage[numbers]{natbib}
\usepackage{tikz}
\usepackage{amsmath}
\usepackage{breakurl}

\usepackage{tikz}
\newcommand*\emptycirc[1][1ex]{\tikz\draw (0,0) circle (#1);} 
\newcommand*\leftcirc[1][1ex]{%
	\begin{tikzpicture}
	\draw[fill] (0,0)-- (90:#1) arc (90:270:#1) -- cycle ;
	\draw (0,0) circle (#1);
	\end{tikzpicture}}
\newcommand*\fullcirc[1][1ex]{\tikz\fill (0,0) circle (#1);} 

\usepackage{authblk}
\usepackage{lipsum}

\begin{document}


\title{\Large \bf \textsc{Donapi}: Malicious NPM Packages Detector using Behavior Sequence Knowledge Mapping}


\author[1]{Cheng Huang}
\author[1\dag]{Nannan Wang}
\author[1]{Ziyan Wang}
\author[1]{Siqi Sun}
\author[1]{Lingzi Li}
\author[1]{\\Junren Chen}
\author[1]{Qianchong Zhao}
\author[1]{Jiaxuan Han}
\author[1]{Zhen Yang}
\author[2]{Lei Shi}

\affil[1]{\textit{Sichuan University}}
\affil[2]{\textit{Huawei Technologies}}


\maketitle
\newcommand\blfootnote[1]{%
\begingroup
\renewcommand\thefootnote{}\footnote{#1}%
\addtocounter{footnote}{-1}%
\endgroup
}
\pagestyle{empty}
\begin{abstract}
With the growing popularity of modularity in software development comes the rise of package managers and language ecosystems. Among them, npm stands out as the most extensive package manager, hosting more than 2 million third-party open-source packages that greatly simplify the process of building code. However, this openness also brings security risks, as evidenced by numerous package poisoning incidents.

In this paper, we synchronize a local package cache containing more than 3.4 million packages in near real-time to give us access to more package code details. Further, we perform manual inspection and API call sequence analysis on packages collected from public datasets and security reports to build a hierarchical classification framework and behavioral knowledge base covering different sensitive behaviors. In addition, we propose the \textsc{Donapi}, an automatic malicious npm packages detector that combines static and dynamic analysis. It makes preliminary judgments on the degree of maliciousness of packages by code reconstruction techniques and static analysis, extracts dynamic API call sequences to confirm and identify obfuscated content that static analysis can not handle alone, and finally tags malicious software packages based on the constructed behavior knowledge base. To date, we have identified and manually confirmed 325 malicious samples and discovered 2 unusual API calls and 246 API call sequences that have not appeared in known samples.
\end{abstract}

\section{Introduction}

JavaScript\blfootnote{\dag \quad Corresponding Author} is increasingly popular as the demand for web applications and web-based applets grows. It has become one of the most widely used programming languages. Many developers use third-party open-source libraries \cite{Hope2020Open, QualityClouds2021Open} to ensure development efficiency and reduce costs by leveraging existing solutions. Open-source development models help avoid redundant development efforts and lower expenses. Additionally, statistics \cite{Nikola2020JavaScript, James2012unfortunate} show that almost every JavaScript package relies on dependencies, and a project (e.g., next.js\footnote{https://www.npmjs.com/package/next}) may depend on hundreds of third-party packages.

\textit{Npm} (\textit{node package manager}) is the default package manager for the JavaScript runtime Node.js, consisting of an online repository and a CLI (command-line interface). Since its first introduction in 2009, it has grown in popularity due to its ease of use and substantial package repository, with over 1.3 million packages and 100 billion downloads per month \cite{Dotan2023Review}, surpassing other package managers such as \textit{Maven} and \textit{pip} as the most extensive package manager. While some companies and organizations maintain their package registries for security reasons, npm remains the most widely recognized and used package registry in the JavaScript community.

As npm gains popularity among developers, it also attracts the attention of attackers. In 2022, an incident \cite{Liran2021war} occurred where RIAEvangelist, the maintainer of the \textit{node-ipc}\footnote{https://www.npmjs.com/package/node-ipc} package, introduced malicious code into the repository and the affected versions \textit{@10.1.1} and \textit{@10.1.2} contained code that overwrote disk files. Subsequently, the attacker created the \textit{peacenotwar}\footnote{https://www.npmjs.com/package/peacenotwar} module, which was included in the affected \textit{node-ipc} version and garnered over 1 million weekly downloads. Another example is the \textit{chalk-next} package, which mimicked a well-known package called \textit{chalk}\footnote{https://www.npmjs.com/package/chalk} that modifies string styles in the console. The attacker used the same \texttt{README.md} to deceive developers and enable information theft. These and others involving packages like \textit{eslint-scope}\footnote{https://www.npmjs.com/package/eslint-scope} and \textit{left-pad}, highlight that poisoning attacks targeting npm are no longer isolated cases but are increasingly common occurrences \cite{ReversingLabs2024Report}.

In this paper, we propose \textsc{Donapi}, which takes sensitive application programming interfaces (APIs) and behavior sequences as the primary research object and combines static and dynamic analysis techniques to detect malicious packages and classify their categories automatically. The primary objective of \textsc{Donapi} is to assist developers in establishing a secure dependency base by facilitating fast package review, reducing the need for manual review, and proactively prevent of the usage of malicious packages. As such, the detector focuses on three key aspects: \textbf{speed}, \textbf{accuracy}, and \textbf{comprehensiveness}. To evaluate its effectiveness, we collected a dataset of over 4,000 publicly available malicious packages and performed several analyses, yielding the following findings:

\begin{itemize}
    \item Batch poisoning: numerous malicious packages often possess similar names while containing identical payloads for executing malicious activities.
    
    \item Attack purposes: malicious packages have various objectives, such as importing malware, stealing information, creating reverse shells, and modifying files.
    
    \item Entry files: the attack payload of a malicious package is commonly embedded in entry files.
    
    \item Code transformation: malicious packages employ code transformation techniques such as obfuscation or compression to evade static detection.
    
\end{itemize}

Based on these findings, we primarily focus the analysis of our detector on the entry files. Research \cite{ohm2020towards} has shown that malicious packages often involve extensive network and file operations. Therefore, we perform static analysis to extract API call sequences from the entry file and its dependent files. These extracted sequences serve as features for training classifiers to make an initial assessment of whether a package is malicious or not. Furthermore, the malicious packages that pass the static screening and obfuscated packages that cannot be handled undergo dynamic analysis. This step aims to extract additional API call sequences and convert them into sequences representing sensitive package behavior. Ultimately, our detector provide hierarchical classification results of behavior sequences, which is not available in the current work \cite{sejfia2022practical, DuanAKESL21, ladisa2023feasibility}.  In summary, the main contributions of our work are as follows:
\begin{itemize}

    \item \textit{Behavior Knowledge Base}: a \textit{hierarchical classification framework} \cite{Donapi2024} based on 806 sensitive API calls and 44 behavior sequences that can automatically map the five most common categories for malicious packages.
    
    \item \textit{Detector \textsc{Donapi}}: an automated malicious package detector, consisting of six primary modules, mixes code analysis, machine learning, and natural language processing techniques to directly map the final malicious category for each detected package.
    
    \item \textit{Effective Results}: we build a local npm package cache of over 3.4 million packages, capable of synchronizing official replicate in near real-time and retaining deleted malicious packages. \textsc{Donapi} found 325 new malicious packages with manual check, discovered 2 unusual API calls and 246 API call sequences that have not appeared in previous malicious samples.
    
\end{itemize}


\section{Background}\label{sec: Background}

From an attacker's perspective, software supply chain attacks consists of three essential steps \cite{sejfia2022practical}: (1) publish a malicious package; (2) get users to install it; (3) run the malicious code. The first two steps are intricately connected, with the success of the second step greatly influenced by the approach employed in the first step. Moreover, these initial two steps form the foundation of a software supply chain attack. Consequently, we can further categorize these three steps into two distinct phases: preparation and execution.

\subsection{Preparation}\label{subsec: Preparation}

While the preparation phase is not the primary focus of our study, we will provide an overview here. One of the simplest methods in this phase is to release a new package. However, attackers often encounter obstacles at this step because of incomplete or suspicious information, which makes users cautious. In this context, attackers employ several methods.

\textbf{Typosquatting} \cite{tschacher2016typosquatting, taylor2020spellbound, neupane2023beyond}. Attackers employ names that closely resemble popular packages when registering names for malicious packages. This deceptive tactic leads some users to download these malicious packages when they make spelling mistakes unintentionally. One notable example is the malicious package \textit{crossenv} \cite{Liran2021typosquatting} mentioned in a Synk blog, which imitates the popular package \textit{cross-env}\footnote{https://www.npmjs.com/package/cross-env}. Despite having similar functionality, the malicious version also includes the unauthorized collection of user information.

\textbf{Dependency Confusion} \cite{Liran2021confusion}. This technique, known as dependency repository hijacking, often employs strategies similar to \textit{Namespacing} \cite{Aviad2021Namespacing}. It involves uploading a package with the same name as certain packages in a private registry but with a higher version number to the public npm registry. As a result, when users synchronize their packages, they unknowingly download the malicious package from a public source. For instance, during one test attack, Birsan \cite{birsan2021dependency} discovered the presence of a package called \textit{auth-PayPal} being used for PayPal development, even though it does not exist in the official npm registry at that time.

On the other hand, a more challenging approach for an attacker is to contribute directly as a maintainer of a legitimate project. In this case, the attacker would then have to gain the privileges needed to modify the code base of the legitimate project by means that typically include using Social Engineering (SE) techniques on legitimate project maintainers \cite{giovanini2021leveraging}, taking over legitimate accounts (e.g., reusing compromised credentials \cite{strongpasswordTute} and account privilege transfers \cite{eventstreamDanny}), 
by compromising the maintainer system (e.g., exploiting vulnerabilities \cite{gokkaya2023software}), and also by exploiting vulnerable points in code maintenance (e.g., outdated domain names of maintainers or lack of package maintenance \cite{zimmermann2019small, Gu2023Investigating}). Finally, they distribute malicious packages that cause users to introduce malicious code (both explicit and implicit) when downloading or updating packages.

\subsection{Execution}
\label{subsec: Execution}

Before the user can execute it, a package must go through two processes, installation and import, which is the stage where the Open Source Security Foundation (OpenSSF)\footnote{https://openssf.org/} has found frequent security problems \cite{Caleb2022Introducing}. A package in npm consists of the \texttt{package.json}\footnote{https://docs.npmjs.com/cli/v9/configuring-npm/package-json} file and various code files that will play a significant role in the above process.

\begin{figure}[!htpb]
    \centering
    \includegraphics[width=\linewidth]{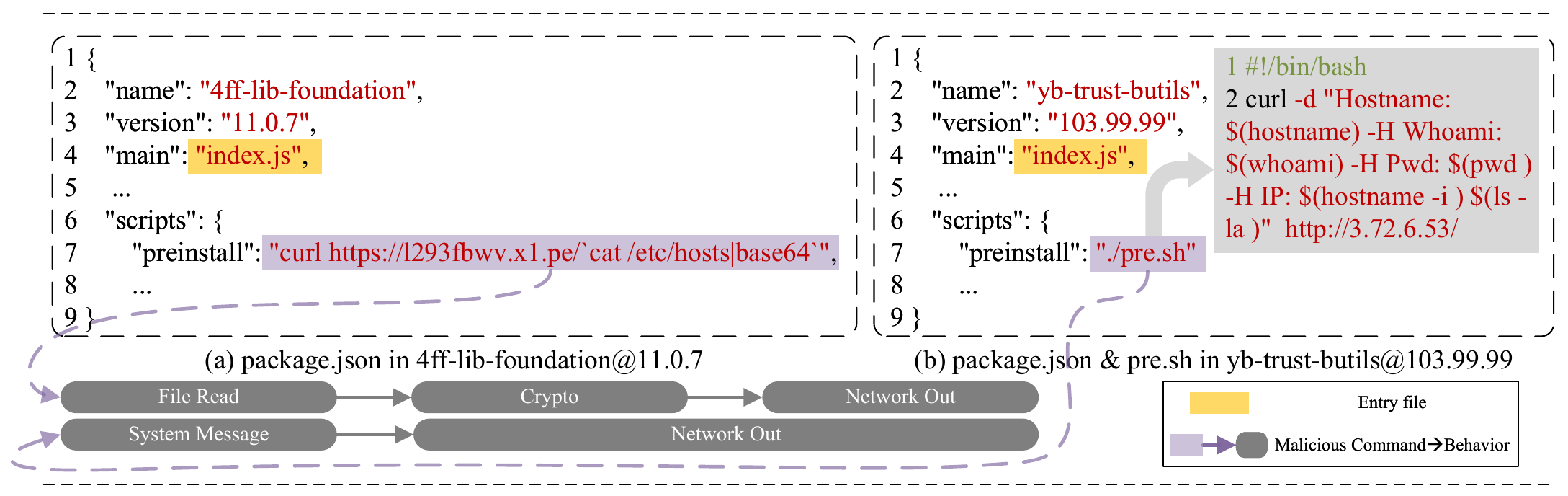}
    \caption{The malicious commands during installation}
    \label{fig: motivation1}
    \vspace{-0.5em}
\end{figure}

\textbf{Installation.} The \texttt{package.json} file involved in the installation process includes a field called \texttt{scripts}\footnote{https://docs.npmjs.com/cli/v9/using-npm/scripts} that provides \texttt{pre} and \texttt{post} hooks for tasks such as preparation and cleanup. While most of these commands are under the user's control, specific fields, such as \texttt{preinstall}, pose a significant security risk \cite{zahan2022weak}. As the name suggests, the commands defined under this field are executed automatically before the users install the package, allowing attackers to launch malicious attacks. As shown in Figure \ref{fig: motivation1}, the package \textit{4ff-lib-foundation@11.0.7} uses a combination of commands in the \texttt{preinstall} field to encode the contents of the local file \texttt{/etc/host} and send it to the external address; the package \textit{yb-trust-butils@103.99.99} calls the \texttt{pre.sh} file in the \texttt{preinstall} and sends the sensitive information to the external address with the \texttt{curl} command.

\begin{figure}[!htpb]
    \centering
    \includegraphics[width=\linewidth]{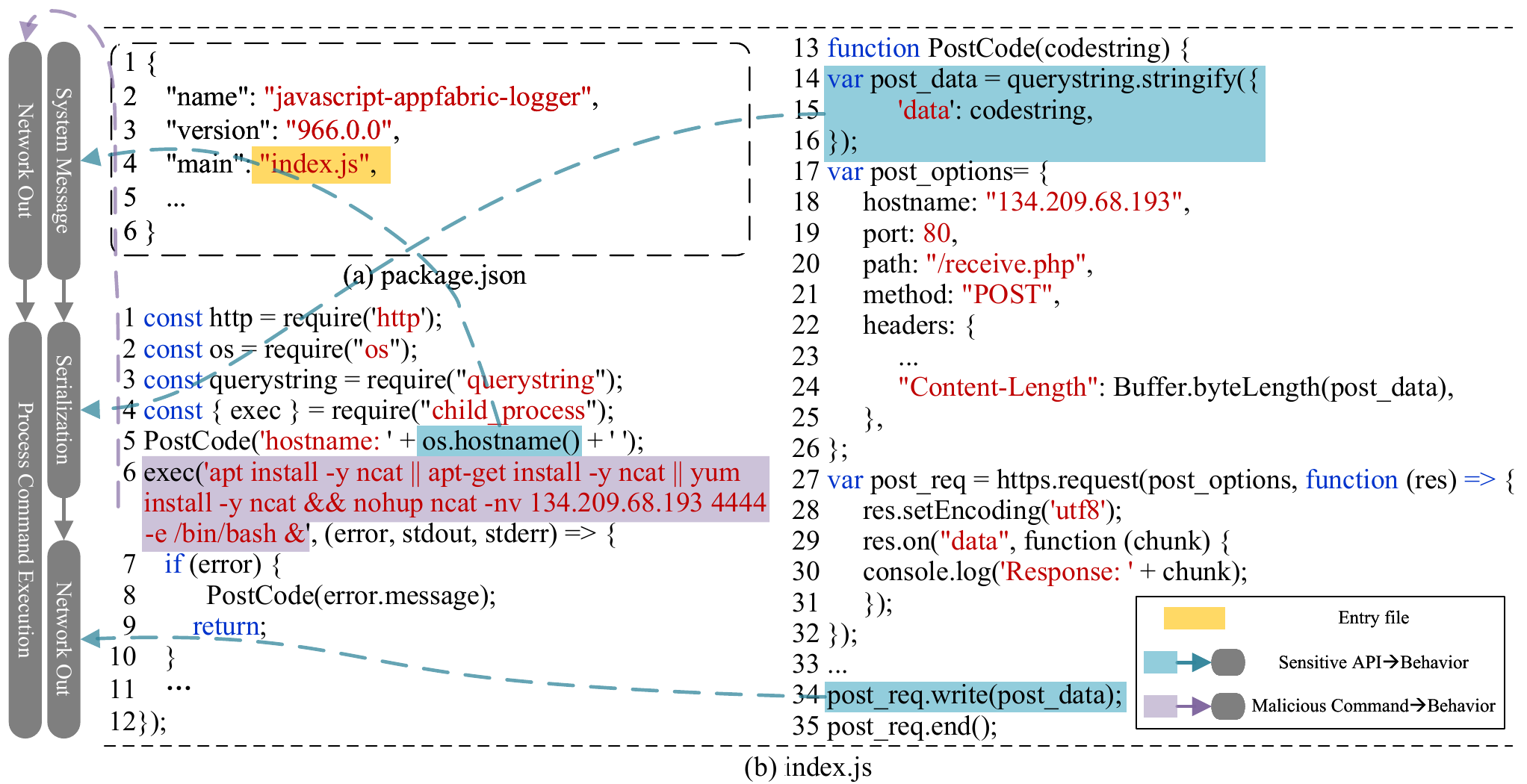}
    \caption{The malicious code in \textbf{javascript-appfabric-logger@966.0.0} detected by \textsc{Donapi} on June 5, 2023. }
    \vspace{-1em}
    \label{fig: motivation2}
\end{figure}

\textbf{Import.} In Node.js, when an object in one code file needs to be accessed by another, it is often necessary to specify it as an exported object. However, some packages import without providing specific export content still have some side effects, such as setting global configurations or executing initialization code. In addition, when the user imports a new module, it automatically executes the file in the \texttt{main} (or \texttt{exports}) field. Therefore, these files is important to consider when detecting malicious packages. As depicted in Figure \ref{fig: motivation2}, if another package imports the package \textit{javascript-appfabric-logger@966.0.0}, it will call multiple APIs to obtain and send sensitive information to an external address.

\textbf{Code transformation techniques.} Code transformation techniques, called code obfuscation in our research scenario, can convert a source program into a target program that is difficult to detect for static analysis without losing its behavior or functionality \cite{DuanAKESL21}. Thus, attackers often use these techniques, including renaming, dead code insertion, control flow flattening, and encoding obfuscation \cite{ren2023empirical}, to mask their malicious intent or increase the analysis complexity. However, it is essential to note that these techniques are not limited to attackers alone. Many legitimate developers also use code transformation techniques to reduce code size or protect their privacy and intellectual property \cite{moog2021statically}. Therefore, obfuscated code alone is not a definitive indicator of malicious intent.

\section{Methods}
\label{sec:Methods}

\begin{figure*}[!htpb]
    \centering
    \includegraphics[width=\linewidth]{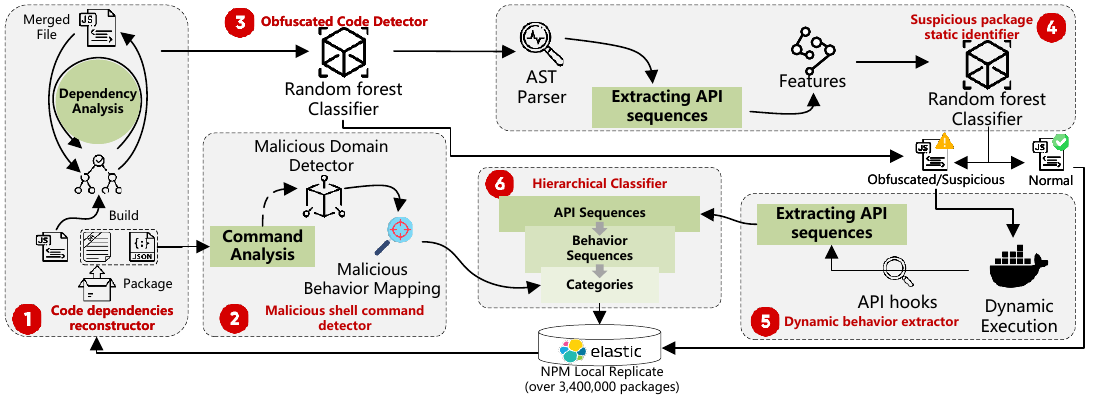}
    \caption{The overall framework of \textsc{Donapi}}
    \label{fig:overview}
    \vspace{-0.5em}
\end{figure*}

In this section, we present the entire processing flow of the detector. The primary labeling technique of the proposed detector is hierarchical classification using behavior sequences derived from API call sequences of the packages. As shown in the Figure \ref{fig:overview}, diagrams 1-6 represent different modules, which we will describe in detail in the subsequent sections.

\subsection{Code Dependencies Reconstructor}

To better capture the API call sequences, we designed a code dependency reconstructor for npm packages that simulates the code execution during the processes of installation and import of the packages by utilizing the Abstract Syntax Trees (ASTs) technique, extracting and merging all the code involved into a single \texttt{.js} file, and rename its parameters, functions, and classes. The process is shown in the Algorithm \ref{algo:Code Dependencies Reconstructor}, and the basic steps and key points are described in detail below.

\textbf{Entry files extraction.} As mentioned in section \ref{subsec: Execution}, the installation and import process is crucial for detecting malicious packages, so we focus on the entrances involved in the above process, including \texttt{scripts}, \texttt{main}, \texttt{exports}, \texttt{imports}, \texttt{bin}. Within the \texttt{scripts} field, apart from the previously mentioned hooks (i.e., \texttt{preinstall} and \texttt{postinstall}), developers can define custom fields executed via the \texttt{npm run <field>} command. We use regular expressions to extract filenames specified under the script field to capture these fields. The \texttt{main} (or \texttt{exports}) field, which serves as the default entry point, automatically executes the appropriate files during package import. The \texttt{imports} field defines the imported subpaths for the current package. \texttt{bin} field creates some CLIs that, in turn, execute specific files. We extract them from the configuration file whenever present, regardless of whether the package executes the specified file upon import.

\textbf{Dependencies parsing.} The purpose of dependencies parsing is to resolve imports of other code files declared dynamically in the code and retrieve the content of the imported target files. In Node.js, different module systems have different import methods, which are \texttt{require()} function for \textit{CommonJS}\footnote{https://wiki.commonjs.org/wiki/CommonJS} and \texttt{import} statement for \textit{ECMAScript}\footnote{https://www.ecma-international.org/publications-and-standards/standards/ecma-262/} modules. Thus, we design two AST node parsing rule sets to adapt to these module specifications. Although a package may contain two module systems, typically, only one module system is used per file, so we generated ASTs in file units. In the AST, each code block (such as function declaration and class declaration) and code lines not contained in code blocks are defined as top-level nodes. The dependencies parsing process traverses the top-level nodes, matches all nodes, including import action, parses them recursively, and inserts the returned AST into the original positions.

\begin{algorithm}[!htb]\footnotesize
    \SetAlgoLined 
	\caption{Code Dependencies Reconstructor}
	\label{algo:Code Dependencies Reconstructor}
	\SetKwInOut{KwResult}{Hyperparameters}
	\KwIn{Package (package.json, code files)}
	\KwOut{Merged codes}
	\KwResult{recursion = 2}
	\SetKwFunction{ExtractEntries}{ExtractEntries}
	\SetKwFunction{GetModuleType}{GetModuleType}
	\SetKwFunction{GenerateAST}{GenerateAST}
	\SetKwFunction{GetDependency}{GetDependency}
	\SetKwFunction{ResolveDep}{ResolveDep}
	\SetKwFunction{RestoreCodeFromAST}{RestoreCodeFromAST}
	\SetKwFunction{MergeAST}{MergeAST}
	\BlankLine
	\tcp{Parsing package.json to get 5 entry points}
	\textit{entries} = \ExtractEntries(package.json)\;
	\ForEach{entry $E_{i}$ in entries}{
	    
	    \tcp{Determine the type of file module specification}
	    
	    \textit{moduleType} = \GetModuleType($E_i$, \textit{package.json})\;
	    
	    \tcp{Generating abstract syntax trees using parsers}
    	$AST_i$ = \GenerateAST($E_i$, \textit{moduleType})\;
    	
    	\ForEach{top-level node $N_{j}$ in $AST_i$}{
    	
        	\If{$N_{j}$ is require/import statement}{
            	\tcp{Get dependencies and download}
        	    \textit{dependencies} = \GetDependency($N_j$)\;
        	    \tcp{Recursive resolving of dependencies}
        	    $AST_{sub}$=\ResolveDep(\textit{dependencies}, \textit{recursion})\;
        	    \MergeAST($AST_i$, $AST_{sub}$)
                
        	    
        	}
        	\ElseIf{$N_{j}$ is exports/export statement}{
        	    \tcp{Modify export statement}
            	variableName = variable name of $N_j$\;
        	    fileName = file name of $E_i$\;
        	    remove export keyword in $N_j$\;
        	    \textit{variableName} $\rightarrow$ \textit{export\_fileName\_variableName}\;
        	}
    	
    	}
    	\tcp{Convert the modified AST back to JS code}
    	$restoreE_i$ = \RestoreCodeFromAST($AST_i$)\;
    	allEntry.append($restoreE_i$)\;
	}
	return allEntry\;
\end{algorithm}

\textbf{Objects modifying.} To solve the problem of variable ambiguity after merging different codes, we need to unify the identifiers of imported and exported objects based on dependency resolution. Since the dependency resolution process will merge code from different files, and the merged code does not need to define any exported objects, we first need to modify the exported objects into general object declarations and unify their names as \texttt{export\_<file\_name>\_<object\_name>}, where \texttt{ file\_name} does not contain extensions and special characters are replaced by "\$" to avoid duplication of variable names. On the other hand, the import statement usually assigns the imported external object to a new identifier to call the external function in the subsequent code, so each identifier member corresponds to an external object during the actual code execution. Based on this principle, we rewrite the value assigned to the new identifier as a new object that includes all exported objects defined as its members in the target file.

\subsection{Malicious Shell Command Detector}

In supply chain poisoning attacks, in addition to directly implanting malicious code, executing the attack through malicious shell commands is a typical tactic hackers use. These commands may exist in various places such as \texttt{package.json} files, \texttt{.sh} files, or code. Therefore, we propose a malicious shell command detector to address this issue.

\textbf{Basic detection process.} For the \texttt{package.json} file, our detector parses and extracts all \texttt{scripts} field values as bash commands and parses them to build the corresponding ASTs using \textit{bashlex}\footnote{https://github.com/idank/bashlex} library. For \texttt{.sh} files, the detector executes the file within a Docker sandbox in debug mode using the command \textit{/bin/sh -x}, capturing the executed command sequence. In the case of code, we focus on the parameters of standard API calls used for command execution (including \textit{child\_process.spawn, child\_process.spawnSync, child\_process.exec, and child\_process.execSync}). In a similar process to \texttt{package.json} file, these parameters are treated as bash commands and processed using the same methodology mentioned above. Then, the YARA rules shown in Appendix \ref{Appsubsec:hierarchical} are used to analyze the information extracted from these bash commands and judge whether they are malicious.

\begin{table}[h]
\centering
\caption{Malicious URL features}
\label{table:Malicious URL features}
\resizebox{\linewidth}{!}{
    \begin{threeparttable}
    \begin{tabular}{cl}
    \toprule
    \textbf{URL Fea.} & \textbf{Description}\\
    \midrule
    UF1 & Longest subdomain entropy\\ \hline
    UF2 & Length of the longest subdomain\\ \hline
    UF3 & \% of vowel letters in the longest subdomain  \\ \hline
    UF4 & \% of consonant letters in the longest subdomain name \\ \hline
    UF5 & \% of consecutive letters in the longest subdomain \\ \hline
    UF6 & \% of repeated letters in the longest subdomain \\ \hline
    UF7 & \% of numeric characters in the longest subdomain \\ \hline
    UF8 & Modify file permissions and create processes \\ \hline
    UF9 & gibberish test to determine the readability of the longest subdomain \\ \hline
    UF10 & Top-level domain types, including: xyz, br, us, etc. \\
    \bottomrule
    \end{tabular}
    \end{threeparttable}
}
\vspace{-0.5em}
\end{table}

\textbf{Malicious URL detection process.} In practical scenarios, specific commands rely solely on accessing URLs to fulfill their intended functions, which means that the degree of maliciousness of the URLs will determine whether the command is malicious. With the attacker's ability improvement \cite{Balaji2023Finding}, the similarity between malicious and benign URLs among text features has dramatically increased, challenging the static detection methods of malicious URLs based on text features. At present, a feasible way is to access the target URL in a sandbox environment, capture the traffic during the communication, and analyze the maliciousness of the URL on this basis. However, this method is time-consuming because the URL needs to be accessed, which makes it challenging to meet the detection efficiency requirement of \textsc{Donapi}. Therefore, we consider a compromise: allowlist combined with machine learning. Specifically, first, we build a domain allowlist based on Alexa Top 1M and use it to filter the target URLs. Next, if the URL fails to hit any items in the allowlist, a machine learning model is introduced to detect it to balance accuracy and efficiency. Regarding machine learning models, referring to existing works \cite{li2020improving, chen2020intelligent}, we use features shown in Table \ref{table:Malicious URL features}.

After completing the above detection process and obtaining the results, we record all malicious commands and forward the packages without malicious commands to the downstream tasks for further analysis.

\subsection{Obfuscated Code Detector}
\label{sec:Obfuscatedcodedetector}

As described in Section \ref{subsec: Execution}, code obfuscation affects the effectiveness of static analysis, so we designed a machine-learning-based model specifically to identify packages with obfuscated code during installation and import. Ultimately, we send obfuscated packages to the \textit{dynamic behavior extractor} and unobfuscated to the \textit{suspicious package static identifiers}.

\begin{table}[ht]
\caption{Obfuscation features. "*" in "Ref." indicates that the feature was partially modified in our research}
\label{table:Obfuscation features easy}
\resizebox{\linewidth}{!}{
    \begin{threeparttable}
    \begin{tabular}{clc}
    \toprule
    \textbf{Obfuscation Fea.} & \textbf{Description} & \textbf{Ref.}
    \\
    \midrule
    OF1 & \makecell[l]{The ratio of compressed lines of code to original lines of code} & new \\
    \hline
    OF2 & \makecell[l]{The ratio of the number of spaces in the compressed code to \\ the number of spaces in the original code} & new \\
    \hline
    OF3 & \# of string function calls, e.g. "subString", "charAt" & \cite{alazab2022detection}\\
    \hline
    OF4 & \# of encoding function calls, such as "escape", "String" & new \\    \hline
    OF5 & \# of occurrences of special characters, such as "\%", "\$", "$\backslash$" & \cite{tellenbach2016detecting}* \\    \hline
    OF6 & \# of lines of code & \cite{likarish2009obfuscated} \\    \hline
    OF7 & \# of white spaces & \cite{likarish2009obfuscated} \\    \hline
    OF8 & \# of special numbers, such as hex and unicode encoding & \cite{likarish2009obfuscated} \\    \hline
    OF9 & Ave. length of identifiers & new\\    \hline
    OF10 & Shannon entropy of identifiers & \cite{kim2011suspicious}* \\
    \hline
    OF11 & Max. string length & \cite{he2018malicious} \\    \hline
    OF12 & \# of strings over a certain length & new \\    \hline
    OF13 & \# of prototype method calls & new \\    \hline
    OF14-OF25 & Frequency of the keyword, such as 'if', 'else' & \cite{tellenbach2016detecting}* \\
    \bottomrule
    \end{tabular}
    \end{threeparttable}
}
\end{table}

Numerous recent advancements \cite{fass2018jast, he2018malicious} have been in obfuscated code detection, but we remain committed to building feature-based detectors due to the two distinct advantages. First, previous research work \cite{kim2011suspicious, likarish2009obfuscated, alazab2022detection,  tellenbach2016detecting,he2018malicious} has extensively used and effectively validated the effectiveness of features in detecting obfuscated code. Second, the feature extraction process can be executed quickly, thus improving overall efficiency. Therefore, based on the study of commonly used obfuscation tools and methods \cite{DefinitiveGuide, ren2023empirical}, we have carefully distilled a set of 25 features based on keywords and code structure and evaluated their importance, as shown in Appendix \ref{Appsec:Obfuscation}. Table \ref{table:Obfuscation features easy} details this comprehensive set of features.

For the model, we chose the Random Forest (RF) model based on several primary reasons. First, RF excels at handling high-dimensional data with many features. Second, in real-world applications with often missing data, RF is just as good at managing missing data while maintaining model accuracy, which may be difficult for other algorithms to produce. Third, RF can help identify the most influential features in distinguishing obfuscated data from unobfuscated data, thereby revealing the underlying mechanisms of code obfuscation. Furthermore, researchers \cite{li2020building, hosseini2019hybrid} have generally favored RF algorithms for detection tasks over other traditional machine learning algorithms, such as Support Vector Machines (SVM), Decision Trees (DT), and Naïve Bayes (NB).

\subsection{Suspicious Package Static Identifier}
\label{sec:Suspiciouspackage}

The main objective of the static detection module is to perform a robust and efficient maliciousness analysis of packages detected by the previous module as unobfuscated files (involved in the process of installation and import). However, considering that static analysis has certain limitations (e.g., poor environment awareness and low precision \cite{Richard2023Static}) compared to dynamic analysis and the combination of them can perform better \cite{Mikhail2023Silent, kang2023scaling}, another goal of this module is to perform preliminary maliciousness assessments on a large number of unobfuscated packages and screen out suspicious samples to be sent to the dynamic module for further examination, to reduce the burden of dynamic analysis and speed up the overall process. The module includes the following two main functional components:

\textbf{API call sequence extraction.} In addition to referring to existing works \cite{DuanAKESL21, sejfia2022practical}, we manually analyze many actual samples and ultimately focus on four different behaviors during script code execution. 1) \textit{Network requests} allow communication over various protocols, such as sockets, HTTP, FTP, etc. They are often used to leak sensitive information \cite{Ionut2023Dozens}, obtain malicious payloads \cite{Catalin2020Malicious}, etc. 2) \textit{File system accesses} allow file operations such as read, modify, chmod, etc. They have been used to compromise SSH private keys \cite{Ionut2023Dozens}, overwrite files \cite{Liran2021war}, etc. 3) \textit{Process operations} allow process creation, termination, and privilege changes. They have been used to spawn individual malicious processes \cite{Jonathan2022Common}. 4) \textit{Arbitrary code execution} allows code generation and execution. The infamous \texttt{eval} implements almost all possible functions.

To capture these behaviors as much as possible, we use Node.js native APIs that implement the four behaviors described above in the underlying layers of Node.js. Additionally, even if executed through dependencies or wrapper functions, the code reconstructor can merge code into a single file for static analysis, meaning that this approach does not miss sensitive behaviors. Drawing parallels to taint analysis \cite{wang2023taintmini, kang2023scaling}, the foundation of our static analysis relies on API call sequences, which aims to ascertain if there is a data flow from source to sink. However, due to efficiency constraints inherent in taint analysis, we opt to utilize solely the location information of API calls to establish the presence of data flow.

We adhere to the "\textbf{AST first, regex later}" methodology to extract sequences. AST is a potent tool for syntax analysis and identifying pivotal system calls. However, it is essential to acknowledge that the diverse nature of JavaScript syntax and the inherent limitations of parsing tools may impede successful parsing across the board. When conventional parsing encounters obstacles, we resort to regex matching as a viable alternative. While regex exhibits more constraints than AST, it proves satisfactory in this context, especially considering its role as a fallback option.

\textbf{Feature engineering.} Feature selection plays a key role, as we initially chose the RF model to identify malicious packages. Striking the right balance is essential: if the feature selection is overly strict, it may lead to a substantial number of missed detections, thereby impacting the effectiveness of subsequent analysis; conversely, if the feature selection is too lenient, it can result in a high number of false positives, significantly impeding the overall efficiency of the \textsc{Donapi}.

\begin{table}[h]
\vspace{-0.4em}
\centering
\caption{Behavior features}
\label{table:Behavioral features easy}
\resizebox{\linewidth}{!}{
    \begin{threeparttable}
    \begin{tabular}{cl}
    \toprule
    \textbf{Behavior Fea.} & \textbf{Description} \\
    \midrule
    BF1 & Send sensitive information to the outside  \\
    \hline
    BF2 & Query system environment variables\\
    \hline
    BF3 & Download the content and execute it as a string  \\
    \hline
    BF4 & Write to file and execute \\
    \hline
    BF5 & Read the contents of the file and execute \\
    \hline
    BF6 & Read files and execute code dynamically \\
    \hline
    BF7 & Download content and execute code dynamically \\
    \hline
    BF8 & Modify file permissions and create processes \\
    \hline
    BF9 & Identify operating system platforms \\
    \hline
    BF10 & Modify the data flow of system command execution results \\
    \hline
    BF11 & Execute system commands \\
    \hline
    BF12 & Number of Performing sensitive file operations \\
    \bottomrule
    \end{tabular}
    \end{threeparttable}
}
\vspace{-0.6em}
\end{table}

Through exploration, we found that identifying malicious behaviors only by individual suspicious behaviors would generate high false positives and decrease the overall analysis efficiency. So, we chose to detect potentially malicious behaviors during the installation phase of software packages through a combination of behaviors as selected features. In addition, existing studies have found significant differences between benign and malicious packages in performing critical behaviors, mainly including a range of native APIs performing essential functions such as file system access, network requests, process manipulation, and arbitrary code execution. For example, when stealing sensitive information, the code execution behavior manifests as a network request and attempts to access sensitive information before that step. In addition, we are interested in modules such as the \emph{fs} module, the \emph{https} module, and the \emph{child\_process} module. Ultimately, we summarized 12 features in Table \ref{table:Behavioral features easy}.

\subsection{Dynamic Behavior Extractor}
\label{subsec:Dynamic behavior extractor}

As described in Sections \ref{sec:Obfuscatedcodedetector} and \ref{sec:Suspiciouspackage}, dynamic analysis helps to overcome the challenges encountered in static analysis, especially in encryption, obfuscation and compression in JavaScript code. Therefore, this module aims to perform runtime API monitoring with API call sequence extraction for the obfuscated and suspicious packages in the previous sections.

\textbf{Monitoring methods.} Most of the existing research \cite{Razgallah2021Behavioral, DeepaK2019IoAm} on capturing API call sequences uses system call tracing tools  (e.g., Strace\footnote{https://github.com/strace/strace}, Sysdig\footnote{https://github.com/draios/sysdig}), which are simple and effective, but lack interpretability. Based on the open-source nature of Node.js, we use API instrumentation that avoids capturing non-package behaviors while having broader API coverage (especially APIs that do not result in system calls). Specifically, we insert additional functional JavaScript code into key API implementations to capture API calls and explicit parameters generated by packages during installation and import. Moreover, we can locate the APIs called by the package to specific code snippets based on call stack information, thus allowing for an intuitive understanding of the package logic at the code level and improving interpretability. For the deployment, we engineered a Docker image that recompiles and integrates modified Node.js source code atop the foundational Ubuntu image. Then, we generate separate Docker containers for each package and perform the processes of installation and import within them, closely mimicking the behavior of actual packages and capturing API call sequences precisely.

\textbf{Monitoring scope.} We monitored 132 APIs, including native implementations such as file manipulation, network connection, and process creation. Our analysis suggests these APIs cover all potential API calls that facilitate malicious operations. However, due to the low-level nature of these APIs, package installation will result in some non-user \textit{side effects}. For example, the \textit{https.request} is implicitly called to download package dependencies and file manipulation APIs are used to log the installation process. To address this issue, we have filtered out these extraneous behaviors by analyzing the function call stack of the API and determining the location of parent function calls and function parameters. Our strategy monitors malicious behaviors at the package installation and import stages. Our research indicates that over 90\% of malicious packages activate during these stages (rather than during exported function invocation), so we only considered the execution of code involved in these phases. In our future work, we aim to continually iterate the functional code of API instrumentation and customize the API return values for different scenarios. This strategy will enable us to emulate a range of execution conditions and external dependencies, thereby improving code coverage.

\subsection{Hierarchical Classifier}
\label{subsec:Hierarchical Classifier}

\begin{figure}[!htpb]
    \centering
    \includegraphics[width=\linewidth]{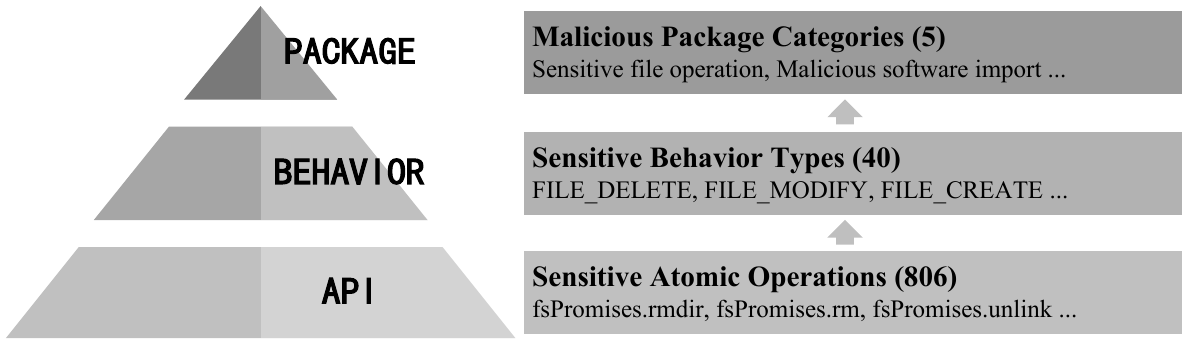}
    \caption{Hierarchical classification framework of malicious packages}
    \label{fig:Hierarchical classification framework of malicious packages}
    \vspace{-0.5em}
\end{figure}

API calls are critical for packages to achieve specific behaviors (benign or malicious) and are the basis of many existing malware classification studies \cite{li2022novel, chen2022cruparamer, amer2021multi}. However, considering that the implementation of package behaviors is not only related to some specific APIs but also greatly depends on the call order of these APIs, e.g., the execution of a malicious software import, which first requires a connection to an external network to download the software, and subsequently to execute the corresponding commands, which can lead to a large number of false positives if the order is not taken into account as can be demonstrated in Appendix \ref{Appsubsec:importance}. Therefore, to enhance the understanding of malicious package behaviors and complement the insufficiency of using API call sets alone, we chose to take the API calls order into account, and we propose a \textit{hierarchical classification framework} centered on the API call sequences, as shown in Figure \ref{fig:Hierarchical classification framework of malicious packages}.

First, we map dynamically extracted APIs to specific behaviors and subsequently classify packages based on the forward and backward order of these behaviors (API call sequence order), using a bottom-up abstraction approach for the whole process. In designing the framework, however, we used a top-down approach to avoid explicitly modeling specific packages based on their characteristics and to be more inclusive. Next, we introduce the three layers of the framework: malicious package categories, sensitive behaviors, and sensitive APIs. For more detailed information, please visit the page \cite{Donapi2024}.

\textbf{Malicious packages.} After studying the existing reports of npm malicious packages \cite{Lindsey2022Dozens, Ionut2023Dozens}, we found that most malicious npm packages steal user information by various means, and a few other types exist \cite{Catalin2020Malicious}. Furthermore, we combed through some of the studies \cite{pachhala2021comprehensive, maniriho2022study} on malware classification, manually analyzed the collected malicious samples, and now classified the malicious packages into five categories. The categories are Sensitive information theft (\textbf{M1}), Sensitive file operation (\textbf{M2}), Malicious software import (\textbf{M3}), Reverse shell (\textbf{M4}), Suspicious command execution (\textbf{M5}). These categories represent different attack purposes despite variations in their specific implementations.

\textbf{Sensitive behaviors.} Malicious packages often use different behaviors and sequences to achieve different purposes. Therefore, we associate the category of malicious packages with the corresponding sequence of behaviors. For example, the sensitive information theft category follows the behavior sequence: "Access sensitive information first and then send it over the network." Therefore, based on the following criteria, we define 12 behavior types (40 behavior subtypes in total).

\begin{itemize}
\item \textbf{Mutual exclusion} (types between atomic behaviors should not overlap)
\item \textbf{Completeness} (covering all possible sensitive behaviors)
\item \textbf{Non-ambiguous} (type division is clear)
\item \textbf{Repeatability} (multiple classification results for an atomic behaviors are consistent)
\item \textbf{Acceptability} (logical and intuitive)
\item \textbf{Practical} (can be used for in-depth research)
\end{itemize}

\textbf{Sensitive APIs.} APIs are fundamental to software functionality and are not inherently malicious. However, depending on the user's different behavioral intentions, certain APIs may contribute to malicious behavior to varying degrees. Therefore, we supplemented the APIs gathered from existing research \cite{DuanAKESL21} with the native API functionality descriptions from the official Node.js documentation\footnote{https://nodejs.org/en/docs}. Ultimately, we identified 226 APIs (combined with parameters to form 806 sensitive APIs) that could potentially be used for malicious purposes to represent the sensitive behaviors defined above at the code level.

The hierarchical classification framework centered on API call sequences is one of our work's highlights, notable for its generality and language independence, allowing for the development of guidelines for different languages. Of course, for consistency, we have also included malicious shell commands in the relevant malicious package category. This additional aspect, while necessary, is separate from the central focus of this section. Therefore, we will present the rules for categorizing malicious shell commands in the Appendix \ref{Appsubsec:hierarchical}.

\section{Experiments and Results}

\subsection{Datasets}
\label{subsec: Datasets}

Our dataset comprises multiple data sources, including security vendor blogs and existing sets of known malicious samples. However, in many cases, these sources only provide information on malicious package names and version numbers without disclosing the specific malicious code itself. Therefore, it poses a challenge when attempting to download packages directly from the official npm registry\footnote{https://registry.npmjs.org/}, which may have been unpublished or deleted, resulting in the unavailability of valid information through the official npm registry.

To overcome this challenge, we implemented a solution by building a local package cache using npm replicate\footnote{https://replicate.npmjs.com/}, which is a CouchDB instance that offers a \textit{Change API}\footnote{https://replicate.npmjs.com/\_changes?descending=true\&include\_docs=true} to track database changes. When building the local cache, we initially copied all the metadata from npm to the local and downloaded all the corresponding \texttt{.tgz} files. Later, we only need to use change messages to perform local synchronization during the synchronization process. However, unlike npm replicate, we only mark packages as officially deleted and do not delete the associated \texttt{.tgz} files. Thus, our dataset includes the raw metadata of the npm registry as well as the \texttt{.tgz} files of the removed packages, some of which may contain malicious packages. We summarize the details of the collected dataset in Table \ref{table:Malicious samples} for reference.

\begin{table}[h]
\centering
\caption{Statistics of the malicious package datasets}
\label{table:Malicious samples}
\resizebox{\linewidth}{!}{
    \begin{tabular}{ccc}
    \toprule
    \textbf{Dataset} & \textbf{Source} & \textbf{Num} \\
    \midrule
    Redlili & \url{https://red-lili.info/} & 1,214 \\
    Backstabber & \url{https://dasfreak.github.io/Backstabbers-Knife-Collection/}  & 1,504 \\
    ReversingLabs & \url{https://blog.reversinglabs.com/blog} & 39 \\
    Maloss & \url{https://github.com/osssanitizer/maloss} & 332 \\
    Cuteboi & \url{https://cuteboi.info/} & 500 \\
    Synk-blog & \url{https://snyk.io/blog/} & 32 \\
    Lofygang & \url{https://gist.github.com/jossef} & 10 \\
    Sonatype-blog & \url{https://blog.sonatype.com/} & 315 \\
    Local cache & - & 600+ \\
    \textbf{Total} & - & \textbf{4,546+} \\
    \textbf{Total (in used)} & - & \textbf{1,159} \\
    \bottomrule
    \end{tabular}
}
\end{table}

We found several problems when studying these datasets: 1) Intersection of data exists; 2) The malicious code is the same except for the external address; 3) The malicious behaviors cannot be triggered (not in installation and import). Therefore, in our analysis, we manually reviewed and de-duplicated these samples and constructed a malicious dataset (1,159 in total). Since some modules of Donapi play complementary roles in detecting malicious packages, this part of the data is highly varied, and we will present it in the corresponding evaluation.

\subsection{Experiment Design}

To evaluate whether our method achieves the three objectives mentioned in the previous sections, we have formulated the following research questions (RQs) to guide the experimental design:

\noindent
\textbf{RQ1 Accuracy.} How does the proposed method perform on the dataset in terms of detection accuracy? (§\ref{subsec:Accuracy evaluation})

\noindent
\textbf{RQ2 Efficiency.}
Can the model handle a large number of packages within a limited time? (§\ref{subsec:Efficiency evaluation})

\noindent
\textbf{RQ3 Validity.}
Can the detector find malicious packages in the wild? How does it compare to other detectors? (§\ref{subsec:Validity evaluation})

\subsection{Accuracy Evaluation (RQ1)}
\label{subsec:Accuracy evaluation}

The \textsc{Donapi} comprises multiple modules, each with specific roles, working collaboratively to achieve malicious package detection. To evaluate the detector, we conducted individual evaluations on each module and assessed the overall performance of the entire detector. The evaluation metrics include precision, recall, and F1 score, considering malicious package detection as a binary classification task. The following sections present each module's experimental setup and results. Furthermore, Section \ref{subsubsec:Integrated evaluation} on integrated evaluation comprehensively explains the causes of false alarms and omissions.

\subsubsection{Partial Evaluation}

\textbf{Code dependencies reconstructor.} We aim to merge the actual running code into a single file while ensuring that the syntax is correct, and we are not concerned with whether or not the merged code will run successfully. In addition, we use the AST for reconstruction, and if there are no errors in the process, we consider the output correct. Since the code merging process is file-based, we use the success rate of the output file as an evaluation metric. We tested all newly released npm packages from May 30, 2023, to June 1, 2023, totaling 28,874 packages and 579,269 files. The overall results, as shown in Table \ref{table:Evaluation results for reconstructor}, demonstrate that the code dependencies reconstructor exhibits excellent usability, achieving an overall error probability of less than 1\%.

\begin{table}[h]
\centering
\caption{Evaluation results for code dependencies reconstructor. "No." is the $n^{th}$ day of the test period}
\label{table:Evaluation results for reconstructor}
\resizebox{\linewidth}{!}{
    \begin{threeparttable}
    \begin{tabular}{ccccc}
    \toprule
    \textbf{No.} &
    \textbf{\#Packages} & 
    \textbf{\#Files} &
    \textbf{\#Reconstruct error files} &
    \textbf{\#Total error files}
    
    \\
    \midrule
    \#1 & 11,917 & 216,615 & 43 & 1,367 (0.63\%)
    \\
    \#2 & 10,481 & 248,901 & 37 & 1,773 (0.71\%) 
    \\
    \#3 & 6,476 & 113,753 & 18 & 1,003 (0.88\%)
    \\
    \textbf{Total} & 28,874 & 579,269 & 98 & 4,143 (0.72\%)
    \\
    \bottomrule
    \end{tabular}
    \begin{tablenotes}    
        \footnotesize               
        \item[1] The total error consisting of Reconstruct errors and Parsing errors.
    \end{tablenotes}
    \end{threeparttable}
}
\vspace{-1.5em}
\end{table}

\begin{table}[ht]
\centering
\caption{Evaluation results for sub-detectors and \textsc{Donapi}}
\label{table:Evaluation results for accuracy}
\resizebox{\linewidth}{!}{
    \begin{threeparttable}
    \begin{tabular}{cccccc}
    \toprule
    \textbf{Detector} & \textbf{\#Malicious/Obfuscated} & \textbf{\#Benign} & \textbf{Prec}. & \textbf{Recall} & \textbf{F1} \\
    \midrule
    MSCD & 208 & 92 & 98.54\% & 97.12\% & 97.82\% \\
    OCD & 88 & 337 & 94.25\% & 93.18\% & 93.71\% \\
    SPSI & 147 & 567 & 99.32\% & 100.00\% & 99.66\% \\

    \textbf{\makecell[c]{\textsc{Donapi}\\(Integral detector)}}  & \textbf{1,159} & \textbf{3,000} & \textbf{98.88\%} & \textbf{91.63\%} & \textbf{95.12\%}  \\
    
    \bottomrule
    \end{tabular}
    \end{threeparttable}
}
\end{table}

\textbf{Sub-detectors.} The main modules for this part of the verification are \textit{Malicious shell command detector (MSCD)}, \textit{Obfuscated code detector (OCD)}, and \textit{Suspicious package static identifier (SPSI)}, which will be replaced with name abbreviations later in the paragraph. As these sub-detectors identify different aspects of a package, we prepared distinct datasets for each of these three components. For \textit{MSCD}, we collected 208 malicious and 92 benign commands, covering the five malicious categories mentioned in Section \ref{subsec:Hierarchical Classifier}, for validation purposes. For \textit{OCD} and \textit{SPSI}, we assembled corresponding datasets of obfuscated and malicious packages and split the training and validation set into a 4:1 ratio. Table \ref{table:Evaluation results for accuracy} shows the specific number of packages and evaluation results. 

\begin{table}[h]
\caption{Comparative experiments on URL classifiers}
\centering
\label{table:URL classifier evaluation}
\resizebox{0.85\linewidth}{!}{
    \begin{threeparttable}
    \begin{tabular}{ccccc}
    \toprule
    \textbf{Model} & \textbf{Acc.} & \textbf{Recall} & \textbf{F1} & \textbf{Speed} \\
    \midrule
    LSTM+CNN+Attention \cite{namgung2021efficient} & 0.91 & 0.68 & 0.79 & \makecell[c]{CPU: 1.465ms/it\\GPU: 0.041ms/it} \\
    Features+RF & 0.82 & 0.58 & 0.62 & CPU: 0.041ms/it \\
    Features+RF+AllowList & 0.82 & 0.58 & 0.62 & CPU: 0.022ms/it \\
    \bottomrule
    \end{tabular}
    \end{threeparttable}
}
\end{table}

In addition, we evaluated the URL classifiers involved in \textit{MSCD} to demonstrate the effectiveness of our design. As shown in Table \ref{table:URL classifier evaluation}, in a CPU-only environment, the machine-learning model, despite being slightly less effective, is much faster than the deep-learning model.  In addition, allowlist can further improve the detection speed without affecting the model's effectiveness.

\textbf{Dynamic behavior extractor.} The Dynamic behavior extractor complements the static identifier and is critical to the overall detection process. We first tested three days of data with a total of 6,766 packages, and 376 were in error (about 5.6\%), and we manually analyzed some error samples. We found that the main reasons for installation failures are code or command errors, such as code syntax errors, no dependency version exists, connection timeouts due to C2 address failures, no installation environment supported, etc. If the dynamic behavior extraction is incomplete for the above reasons, the package cannot successfully attack the victim host despite being malicious. It is worth mentioning that we can still capture network request behavior for packages that fail to install due to connection failures.

\begin{table}[h]
\centering
\caption{Evaluation results for hierarchical classifier}
\label{table:Evaluation results for hierarchical classifier}
\resizebox{0.9\linewidth}{!}{
    \begin{threeparttable}
    \begin{tabular}{ccc}
    \toprule
    \textbf{Module} & \textbf{Category} & \textbf{Recall} \\
    \midrule
    \multirow{5}*{\rotatebox{90}{\makecell{Hierarchical\\ classifier}}}
      & Sensitive information theft (M1) & 93.14\% \\
    ~ & Sensitive file operation (M2) & 100.00\%  \\
    ~ & Malicious software import (M3) & 82.28\% \\
    ~ & Reverse shell (M4) & 97.22\% \\
    ~ & Suspicious command execution (M5) & 68.75\% \\
    \bottomrule
    \end{tabular}
    \end{threeparttable}
}
\end{table}

\textbf{Hierarchical classifier.} To verify the validity of the criteria we defined for our hierarchical classification framework, we used the malware set mentioned in Section \ref{subsec: Datasets}, which covers all five malicious categories we defined. However, it is worth noting that the actual number of categories for packages is slightly more than 1,159 due to the complex behavior of some packages, which results in the possibility of belonging to more than one malicious type, e.g., both sensitive information theft and sensitive file tampering. The specific recall rates for each category are shown in Table \ref{table:Evaluation results for hierarchical classifier}.

We initially designed Suspicious command execution (M5) to encompass more suspicious commands to avoid underreporting. However, after analyzing the missed samples, we found the recall rate unsatisfactory due to their obfuscated code and inability to obtain the complete behavior sequences through dynamic execution. In future work, we will refine the detector based on the characteristics of such samples.

\subsubsection{Integrated Evaluation}
\label{subsubsec:Integrated evaluation}

In the previous sections, we conducted detailed performance evaluations for each module, but assessing the detector as a whole is also essential. The malicious samples used in the experiments come from the data mentioned in Section \ref{subsec: Datasets} (1,159 in Total). For the benign samples, we randomly selected and manually validated 3,000 packages.

In the accuracy evaluation of RQ1, we found that the detection could be better, so we analyzed the samples in detail. For underreporting, we discovered that some malicious samples are time-sensitive, e.g., failed external links cause us to extract API call sequences incompletely. For false positives, we found that the rich behavior of large software packages may satisfy part of our defined API call sequences, which leads to false positives; moreover, the dependencies between packages and between packages and other software lead to the fact that these packages inevitably import external software, but we are unable to determine the degree of maliciousness of the imported software accurately.

\begin{figure*}[htbp]
\centering
\subfloat[Synchronization of local npm package cache per 14 days]{
     \includegraphics[width=0.28\linewidth]{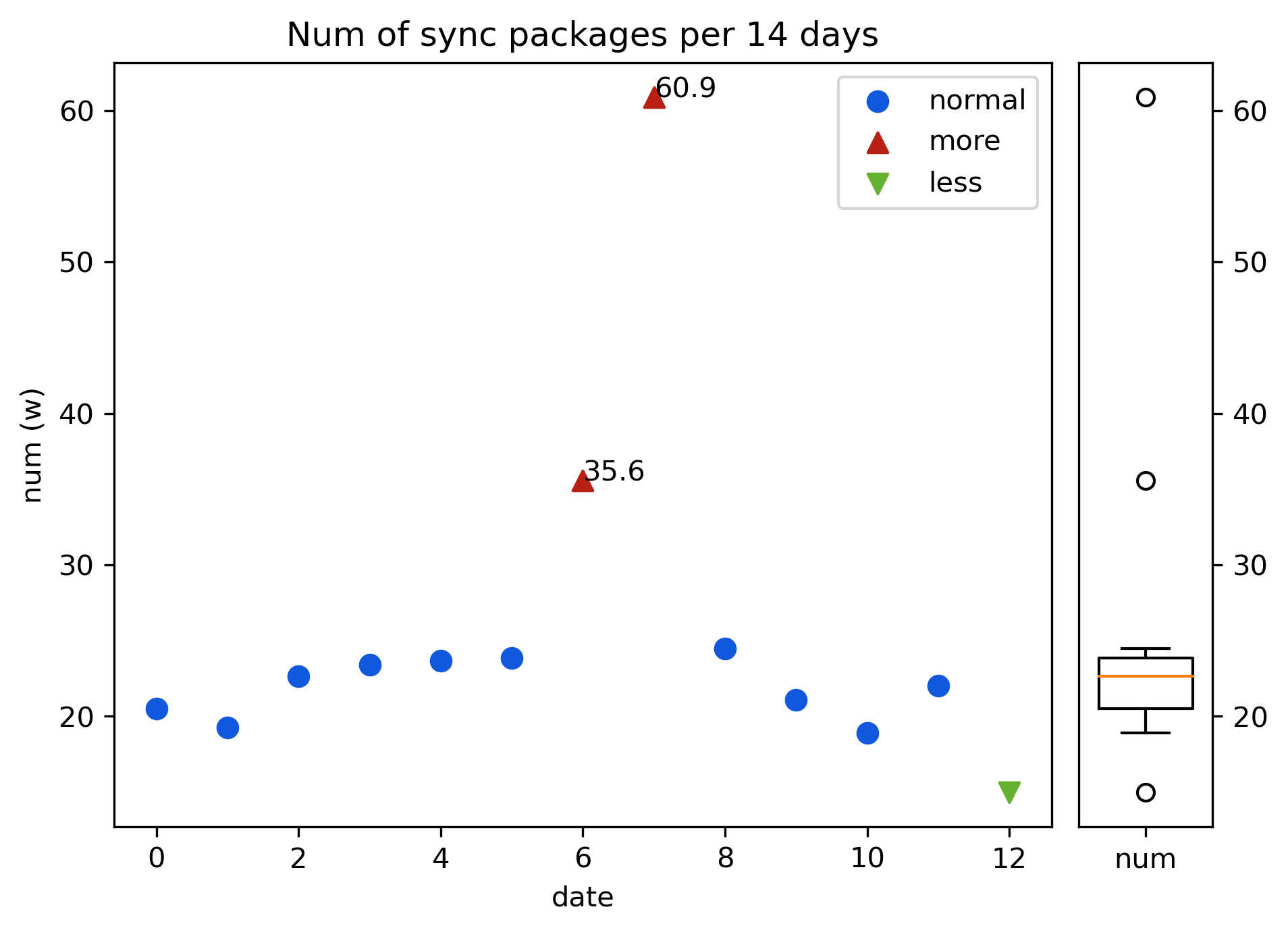}
}
\quad
\subfloat[Synchronization of local npm package cache in May]{
    \includegraphics[width=0.28\linewidth]{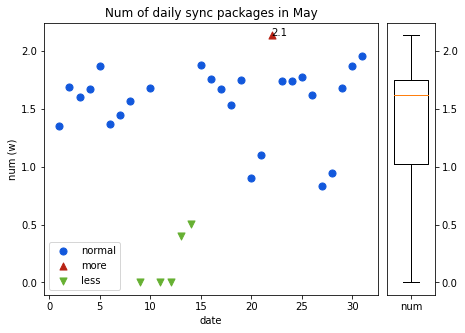}
}
\quad
\quad
\quad
\quad
\quad
\quad
\quad
\quad
\quad
\quad
\quad
\quad
\quad
\quad
\subfloat[eCDF of Static detection runtime]{
     \includegraphics[width=0.30\linewidth]{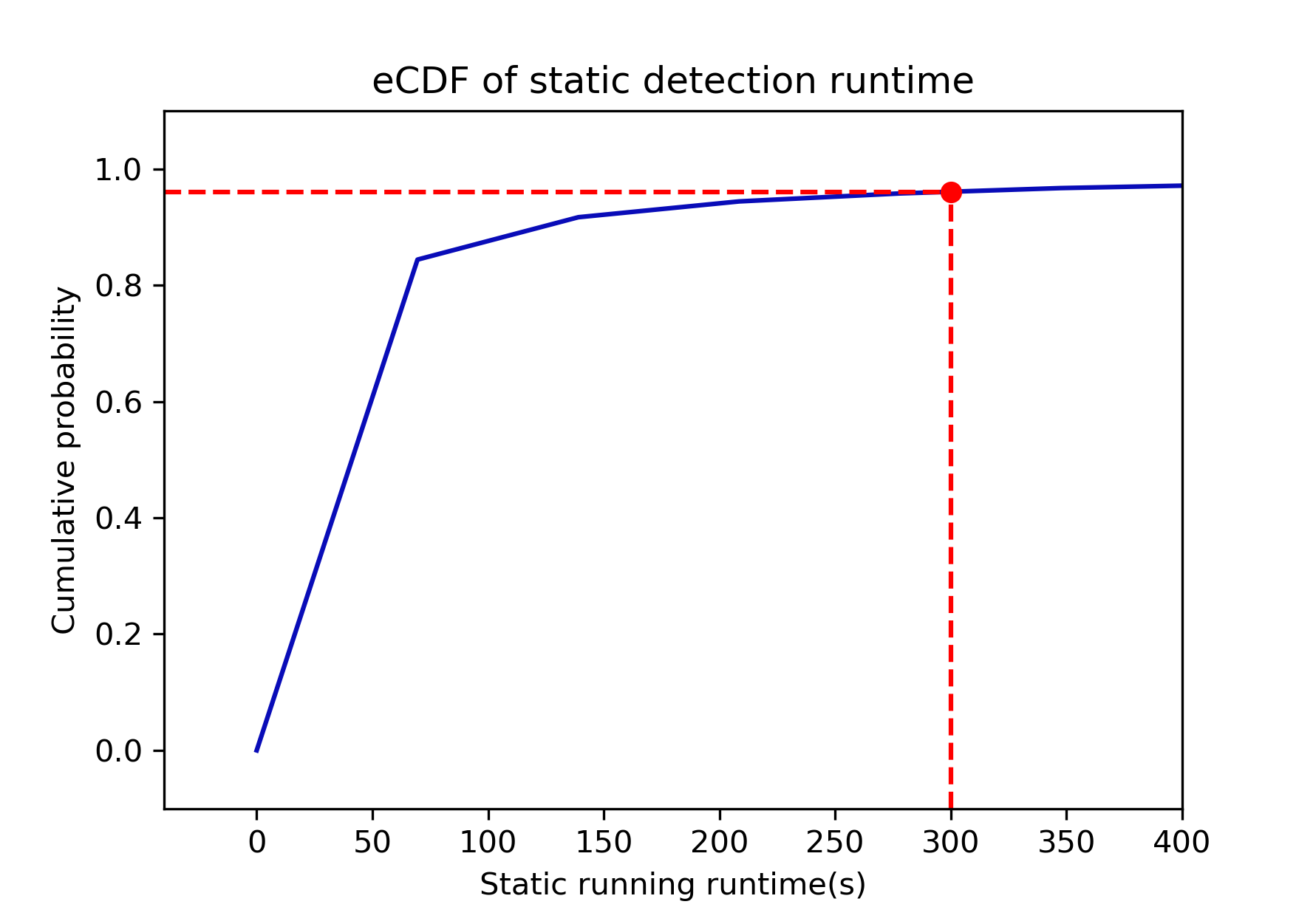}
}
\quad
\subfloat[eCDF of Dynamic detection runtime]{
    \includegraphics[width=0.3\linewidth]{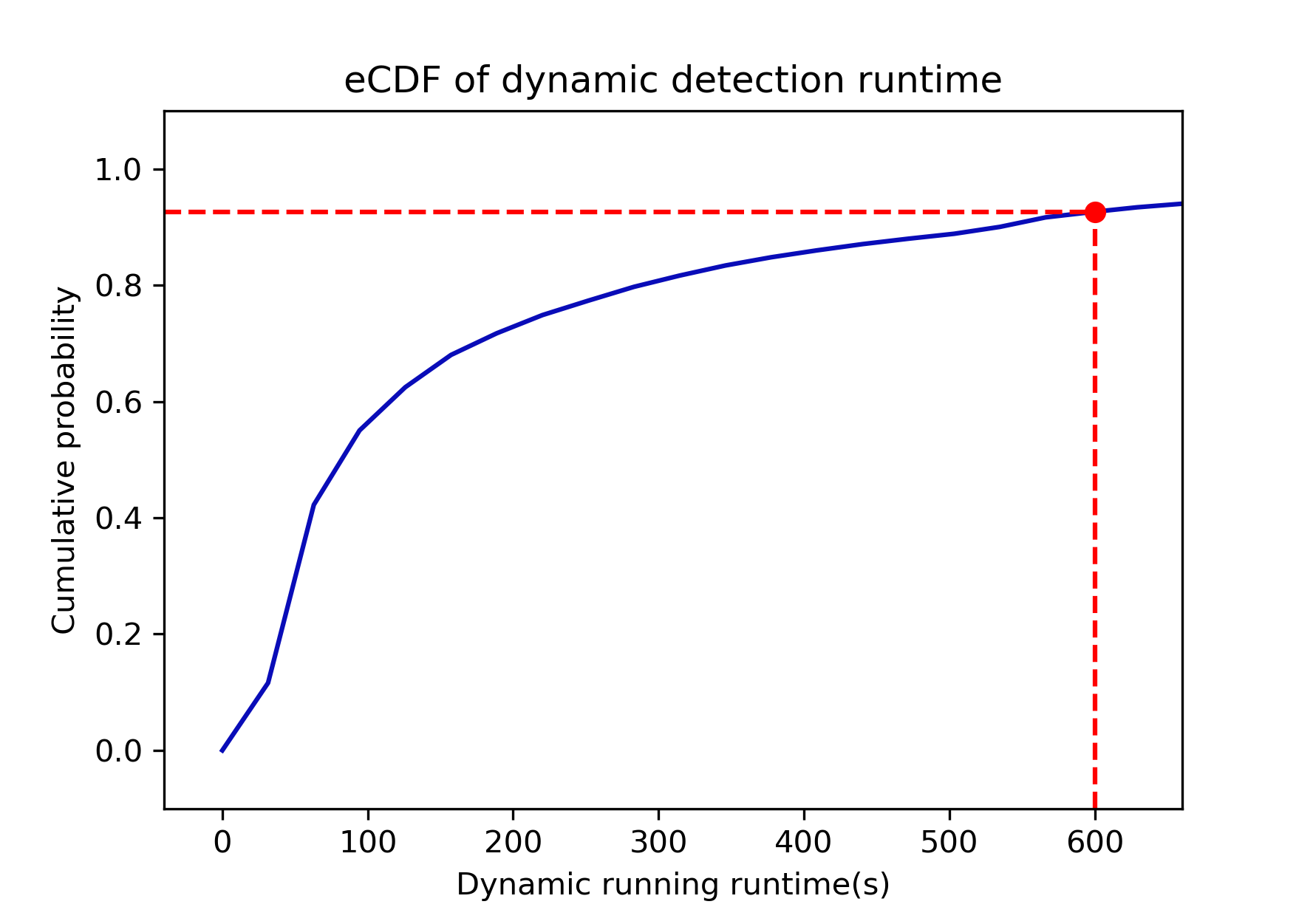}
}
\quad
\subfloat[Distribution of detection runtime]{
    \includegraphics[width=0.3\linewidth]{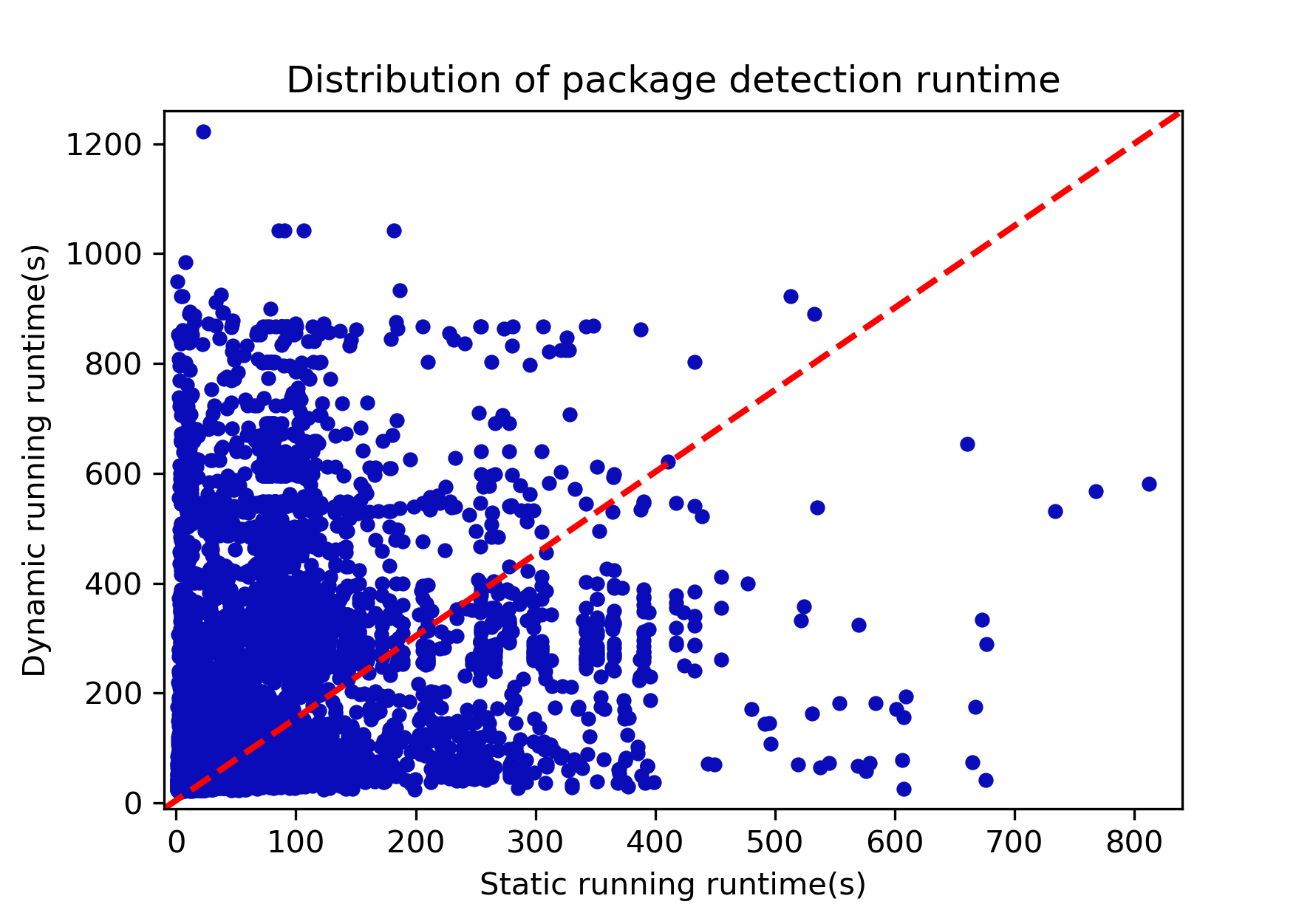}
}
\setlength{\abovecaptionskip}{8pt}
\caption{Evaluation results for efficiency}
\label{fig:Evaluation results for Efficiency}
\vspace{-0.8em}
\end{figure*}

\subsection{Efficiency Evaluation (RQ2)}
\label{subsec:Efficiency evaluation}

Before the efficiency evaluation, we analyzed the update frequency of the npm repository. Figure \ref{fig:Evaluation results for Efficiency}(a) shows the number of package updates in the local npm package cache every 14 days from \textit{December 12, 2022} to \textit{June 24, 2023}. The box plots and marked points represent the distribution of package updates over this period, and we observe that approximately 219,834 packages are released or updated every 14 days, an average of 16,102 package updates per day. Thus, with such many package updates, our detector must operate efficiently without compromising the quality of the analysis.

\subsubsection{Timeout Analysis}
\label{subsubsec:Timeout Analysis}

The dynamic and static API call sequence extraction is a critical detector component and significantly impacts the overall processing time. However, complex package behaviors may lead to excessively long processing times, e.g., large packages may lead to longer static API call sequence extraction times, and network issues may lead to unstable dynamic installation times. Therefore, we employ a timeout mechanism to prevent the detector from taking too long to process specific packages.

To determine suitable timeout values, we conducted tests on a randomly selected subset of packages. For static API call sequence extraction, we tested 50,000 packages and calculated the empirical cumulative distribution function (eCDF) under different time durations, as shown in Figure \ref{fig:Evaluation results for Efficiency}(c). We found that when the detection time was less than 50s, 150s, and 300s, the coverage rate for packages reached 82\%, 93\%, and 97\%, respectively. Based on these results, we set the static processing timeout to 300s to ensure sufficient coverage. However, it is essential to note that most packages require significantly less than 300s for processing.

Similarly, we randomly selected 15,000 packages for dynamic API extraction for testing and analyzed the eCDF under different time durations, as depicted in Figure \ref{fig:Evaluation results for Efficiency}(d). Due to the variations in package functions, there is variability in the duration distribution. Therefore, to achieve comprehensive coverage, a timeout of 600s was set for dynamic execution, as this duration covered nearly 89\% of packages. It is worth mentioning that this does not imply that the processing time is inherently long, as not all packages require dynamic detection.

\subsubsection{Processing Time}
\label{subsubsec:Processing time}

After determining the timeout, we randomly selected actual update packages totaling three days (September 1-3, 2023) for processing duration evaluation. It is worth noting that not all packages go through the dynamic analysis step. Therefore, the experiment tested a total of 15,479 samples, of which 4,571 were processed by the dynamic behavior extractor.

\begin{table}[h]
\centering
\caption{Evaluation results for efficiency. Statistics are from September 1-3}
\label{table:Evaluation results for efficiency}
\resizebox{\linewidth}{!}{
    \begin{threeparttable}
    \begin{tabular}{cc}
    \toprule
    \textbf{Object} & \textbf{Result} \\
    \midrule
    Num of detected packages & 15,479 (4,571 through dynamic)  \\
    Processing time & 21 h 48 m 36s \\
    Total lines of all codes & 168,610,774 rows  \\
    Total lines of reconstruction codes & 19,989,837 rows \\
    \makecell{Num of detected packages in 24 hours \\(estimated)} & $\approx$ 17,033 (> 16,102) \\
    \bottomrule
    \end{tabular}
    \end{threeparttable}
}
\end{table}

Table \ref{table:Evaluation results for efficiency} lists the test results, showing that the detector could process 15,479 packages in 22 hours. Meanwhile, to prevent the number of update packages from fluctuating over time, we estimated the number of packages (about 17,033) processed by the detector over 24 hours, and the results show that it is similar to the average number of updates (16,102) per day that we have counted before. In addition, given the variation in the number of lines per package, measuring the detection efficiency in terms of code lines is necessary. On this basis, the detection efficiency of the detector is about 1.29 million lines per 10 minutes for all codes and about 152,757 lines per 10 minutes for reconstruction code.

\subsection{Validity Evaluation (RQ3)}
\label{subsec:Validity evaluation}

To validate the effectiveness of the detector, we conducted experiments with a larger number of packages and compared its performance with other tools on some packages. This extensive evaluation allows us to understand better the detector's effectiveness in detecting and mitigating malicious packages.

\begin{table}[h]
\centering
\caption{Evaluation results for validity}
\label{table:Evaluation results for validity}
\resizebox{\linewidth}{!}{
    \begin{threeparttable}
    \begin{tabular}{ccccc}
    \toprule
     \textbf{Detector} & \textbf{Term} &  \textbf{Total} & \textbf{Det.} & \textbf{Pos. Det.} \\
    \midrule
    \textsc{Donapi} & Jan-May & 2,764,022 & 1,727 & 325 (+165)\\
    \hline
    \textsc{Donapi} & \multirow{4}{*}{May} & \multirow{4}{*}{420,395} & 792 &  148 (+83) \\
    GUARDDOG \cite{guarddog2022Finding}&                    &                    & 49,070 & $\approx$ 6\,in\,1,000 \\
    AMALFI \cite{sejfia2022practical}&                    &                    & 2,678 &  $\approx$ 22\,in\,1,000\\
    SAP \cite{ladisa2023feasibility}&                    &                    & 50,043 & $\approx$ 6\,in\,1,000\\
    \bottomrule
    \end{tabular}
    \begin{tablenotes}    
        \footnotesize               
        \item[Note] Numbers in parentheses are the number of malicious packets detected by the model but not visually analyzed manually due to code obfuscation.  
      \end{tablenotes}
    \end{threeparttable}
}
\vspace{-0.5em}
\end{table}

\textbf{Long term.} Starting in 2023, we deployed \textsc{Donapi} to real-world environments to detect packages updated daily from our local caches. We identified and manually confirmed 325 malicious packages (and tagged with npm or Synk) from January through June. It is worth noting that the results of our detector needed to be more satisfactory during the early stages of deployment, and the number of samples detected was relatively low. However, as we iterated and improved our detector, by May, our model stabilized, and the detections aligned with our experimental results and expectations. \textsc{Donapi} detected 148 actual malicious packages out of 420,395 packages throughout May. The specific detection results are shown in Table \ref{table:Evaluation results for validity}.

\begin{table}[h]
\centering
\caption{Evaluation results of different tools on datasets}
\label{table:Evaluation results of different tools on dataset}
\resizebox{\linewidth}{!}{
    \begin{threeparttable}
    \begin{tabular}{ccccccc}
    \toprule
     \textbf{Detector} & \textbf{TP} & \textbf{FP} & \textbf{Acc.} & \textbf{Prec.} & \textbf{Recall} & \textbf{F1} \\
    \midrule
    AMALFI \cite{sejfia2022practical} & 1,031 & 27 & 0.97 & 0.97 & 0.89 & 0.97\\
    SAP \cite{ladisa2023feasibility} & 1,083 & 355 & 0.93 & 0.75 & 0.93 & 0.83\\
    GUARDDOG \cite{guarddog2022Finding} & 1,052 & 512 & 0.90 & 0.67 & 0.91 & 0.77\\
    \textsc{Donapi} & 1,062 & 116 & 0.97 & 0.90 & 0.92 & 0.93\\
    \bottomrule
    \end{tabular}
    \end{threeparttable}
}
\vspace{-1em}
\end{table}

\textbf{Comparative study.} In this study, it is necessary to demonstrate the effectiveness of our detector by comparing it with other npm package detection tools. \textit{GUARDDOG}, a heuristic rule-based tool developed by Google, is fully open-source and thus easily used by others for comparison. \textit{AMALFI} and \textit{SAP}, leading approaches based mainly on machine learning, hide part of their code for security reasons, so we reconstructed the feature extraction function according to the paper description. In addition, since \textit{AMALFI} does not provide a training set, we use our dataset for training here, and \textit{SAP} uses the dataset supplied by their repository.

\begin{figure}[h]
    \centering
    \includegraphics[width=0.9\linewidth]{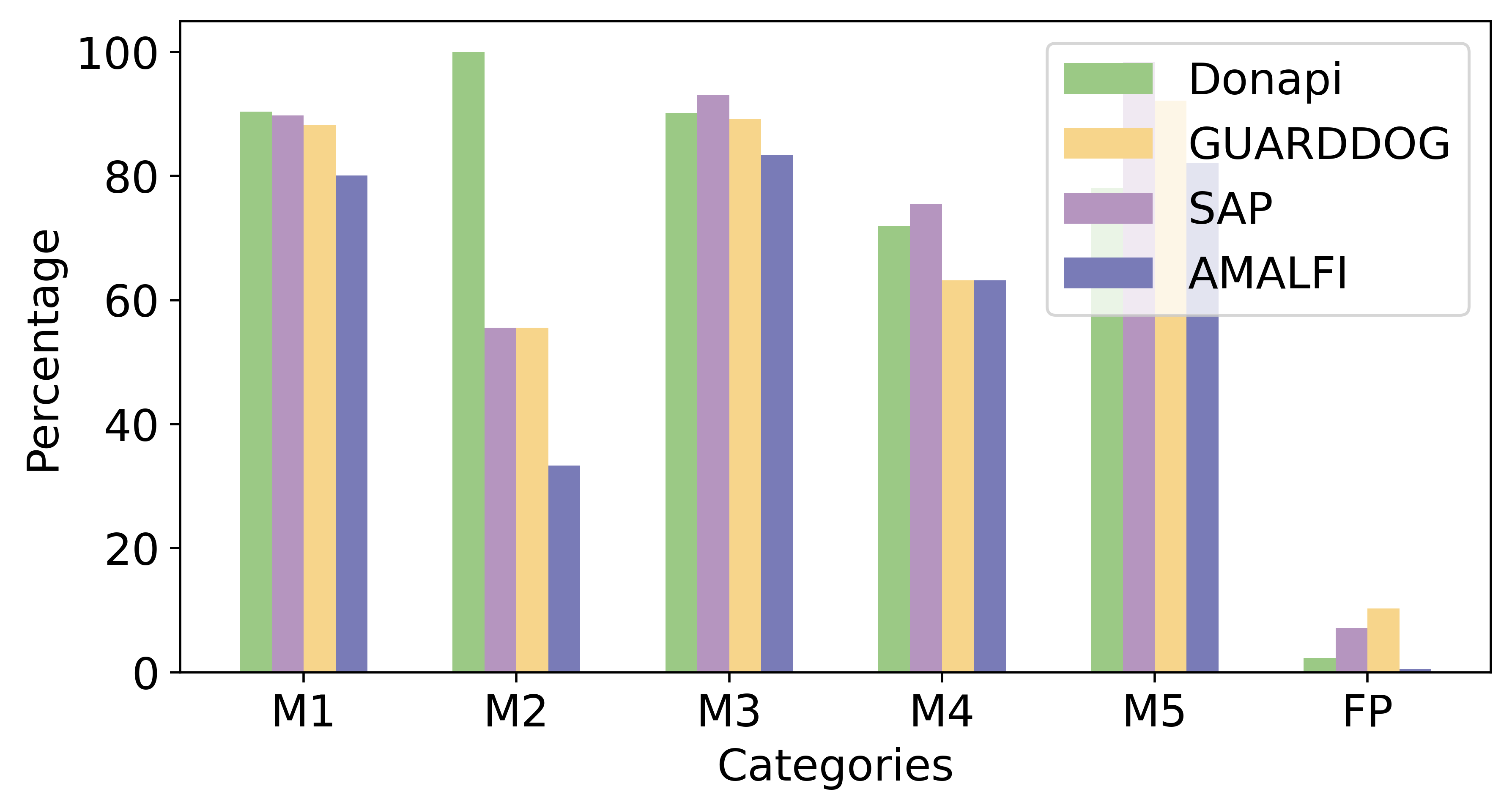}
    \setlength{\abovecaptionskip}{5pt}
    \caption{Percentage coverage of different package categories by different tools}
    \label{fig:valid_eval}
    \vspace{-1.5em}
\end{figure}

We first evaluated them using the malicious samples(1159) used in the Integrate evaluation and the npm Top 5000 packages as benign samples. As shown in Table \ref{table:Evaluation results of different tools on dataset}, \textit{GUARDDOG} and \textit{SAP}'s performance in the false positives and evaluation metrics is significantly worse than that of AMALFI and our detector. Moreover, the approximate detection numbers do not imply similar detection capabilities. As shown in Figure \ref{fig:valid_eval}, different detectors have different detection effects on different malicious categories, and our detector performs more balanced and significantly outperforms the other tools in the M2 category. In addition, we found that out of the 1052 instances detected by \textit{GUARDDOG}, 972 were determined to be malicious based solely on the presence of the \texttt{scripts} field, and \textit{SAP} and \textit{AMALFI} rely heavily on \texttt{package.json}, which contrasts with our API call sequence-based approach and highlights a significant difference in detection methods.

In addition, for better comparison, we also analyzed all packages locally synchronized in May 2023 using these tools, and the results are shown in Table \ref{table:Evaluation results for validity}. The other three tools reported more malicious samples than \texttt{Donapi}, and \textit{GUARDDOG} and \textit{SAP} even reported nearly 50,000. Subsequently, we randomly sampled and manually reviewed the samples reported by the other three tools, as shown in Table \ref{table:Evaluation results for validity}. We found that only 6, 22, and 6 of them are truly malicious samples, respectively, which indicates that our detector has a lower false positive rate in real scenarios, thus reducing the workload of security researchers more effectively.

\section{Discussion}
\label{sec:Discussion}

During our experiments and daily detection, we have gained valuable insights into malicious packages and our detector. Here, we will provide a concise overview of these findings.

\begin{table}[h]
\centering
\caption{Case study of unconventional behavior sequences}
\label{table:Case study}
\resizebox{\linewidth}{!}{
    \begin{threeparttable}
    \begin{tabular}{cccc}
    \toprule
    \textbf{Cat.} & \textbf{Behav.} & \textbf{Conv.} & \textbf{A Case of Unconv.} \\
    \midrule
    \multirow{5}*{\makecell{Sensitive \\information \\theft \\ (M1)}} & 
    FILE\_CREATE & \emptycirc & \leftcirc \\
    ~ & SYSTEM\_MESSAGE & \fullcirc & \fullcirc \\
    ~ & SERIALIZATION & \leftcirc & \emptycirc \\
    ~ & PROCESS\_COMMAND\_EXECUTION & \emptycirc &\leftcirc \\
    ~ & NETWORK\_OUT & \leftcirc & \leftcirc  \\
    
    \hline
    
    \multirow{4}*{\makecell{ Reverse \\shell \\ (M4)}} &
    NETWORK\_IN & \leftcirc & \emptycirc \\
    ~ & PROCESS\_FILE\_EXECUTION & \emptycirc &\leftcirc \\
    ~ & PROCESS\_COMMAND\_EXECUTION & \leftcirc &\leftcirc \\
    ~ & NETWORK\_OUT/IN & \fullcirc & \emptycirc  \\
    \bottomrule
    \end{tabular}
    \begin{tablenotes}    
        \footnotesize               
        \item[1] \emptycirc miss; \t \leftcirc hit once; \t \fullcirc Hit more than once; 
      \end{tablenotes}
    \end{threeparttable}
}
\end{table}

\textbf{Case study.} We will illustrate the advantages of our detector using a few examples from our experiments. During our daily detection operations, we have observed that dynamic execution significantly complements static detection in our detector. This behavior was captured through dynamic execution, providing valuable insights that static analysis alone would have missed. For instance, the packages \textit{businessemailvalidator@99.10.9} and \textit{azure-sdk-v3@99.10.11} employed obfuscation techniques to transform their code, rendering static parsing ineffective. However, our dynamic analysis was still able to capture the behavior of these packages.

Furthermore, our hierarchical classification framework can encompass many malicious samples. For instance, we can accurately classify \textit{@m365-admin/customizations@999.9.15} to M1 and \textit{tslib-tool@1.6.1} to M4, even though their behavior sequences significantly differ from the conventional behavior sequences of our known samples, as indicated in Table \ref{table:Case study}. As an example, the case of M1 sent host information to a remote server by decrypting the A file using AES in the code and executing the malicious code contained within it. Our analysis also discovered two previously unseen APIs: \textit{fs.fchown} and \textit{fs.writeSync}. Thus, using a comprehensive and well-developed list of APIs helps our framework effectively detect these new APIs used in malicious packages.

\textbf{Findings.} After analyzing collected malicious datasets and the malicious samples newly found by our detectors, we found the following phenomena. \textbf{1) Code reuse.} We observed instances where packages released by different authors simultaneously exhibited identical malicious behavior and code, with the only difference being the outgoing network addresses. For example, packages like \textit{smart-jsonapi@1.1.1} and \textit{uitk-build-tasks@10.0.0} showcased this phenomenon. While we cannot definitively conclude that they originated from distinct attackers, it is crucial to acknowledge the prevalence of such cases. Tracing these malicious packages back to the same attacker could imply an association with a specific attack template, enabling us to trace and address the attacking organization more effectively. To facilitate this effort, we are currently compiling a list of payloads from frequently encountered malicious packages, some of which are accessible on our web page. \textbf{2) Sophisticated attack.} As the technology for malicious package detection advances, many malicious packages attempt to bypass detection by collaborating with multiple packages, primarily through dependencies. For instance, in our investigation on the \textit{@alfalab} series, 47 packages were released on a specific day. However, the malicious code was present only in one package named \textit{@alfalab/core-components-spinner}, while the rest of the packages, such as \textit{@alfalab/core-components-action-button} and \textit{@alfalab/core-components-button}, leveraged dependencies to carry out malicious behavior. Similar to the above are packages such as \textit{jpeg-metadata@1.5.1/ttf-metadata@1.5.2} \cite{Phylum2023Sophisticated}, which in their first step obtain a token from one of several potential remote servers, and in their second step use this token to obtain another attack script from a remote server. \textbf{3) 0-days.} During the deployment of \textsc{Donapi}, we found and manually confirmed 325 new malicious packages, all discovered on their release day. We first investigated the inclusion of two vulnerability databases (Synk\footnote{https://security.snyk.io/} and Mend.io\footnote{https://www.mend.io/vulnerability-database/}). We found that nearly 20\% of the packages were not indexed by either, while only 50\% were indexed by both. Then, we investigated how long these malicious packages existed in the npm source and found that they lived for about one week on average, with some of them being able to exist for dozens or even hundreds of days. Therefore, our work can help npm officials find and take down these malicious packages faster.

\textbf{Novelty.} 
Table \ref{table:Existing Tools} summarizes some existing methods for checking the maliciousness of npm packages, compared to which our proposed DONAPI has the following advantages. First, \textsc{Donapi} effectively combines both dynamic and static analysis techniques. Most existing works lack dynamic analysis, which prevents them from effectively analyzing obfuscated packages. Second, \textsc{Donapi} has a more targeted and highly automated detection mechanism. It simulates package installation and import process, analyzes multiple aspects, including command scripts, code, and URLs, and builds a complete knowledge base. Third, \textsc{Donapi} introduces the code dependencies reconstructor. The module simulates the code execution during installation and import, thus avoiding analyzing all files in the package \cite{sejfia2022practical, ladisa2023feasibility}. In addition, it preserves dependency information between files, which is missing when analyzing each file individually \cite{DuanAKESL21}.
Finally, \textsc{Donapi} presents some technical details that previous work overlooked. 1) Improve robustness through anti-evasion measures. (e.g., modifying container environment variables to prevent malicious packages from evading dynamic execution). 2) Using API instrumentation instead of Strace \cite{wyss2022wolf, DuanAKESL21} provides better interpretability by avoiding tracing non-package behaviors and enabling code stalking analysis. 3) Broader API coverage can help to reduce underreporting.

\begin{table}[h]
\caption{Existing tools for analyzing npm packages}
\label{table:Existing Tools}
\resizebox{\linewidth}{!}{
    \begin{threeparttable}
    \begin{tabular}{ccc}
    \toprule
    \textbf{Package scanner} & \textbf{Detection Granularity} & \textbf{Technique used} \\
    \midrule
    npm-audit \cite{NPMaudit2023Package} & Metadata of dependencies & Static (Rules) \\
    Ferreira et al. \cite{ferreira2021containing} & Package  & Static (Rules) \\
    Ohm et al. \cite{ohm2020supporting} & Artifact & Static (ML) \\
    Zahan et al. \cite{zahan2022weak} & Metadata & Static (Rules)\\
    Liang et al. \cite{liang2021malicious} & Package & Static (ML) \\
    SAP \cite{ladisa2023feasibility} & Package & Static (ML) \\
    GUARDDOG \cite{guarddog2022Finding} & Package & Static (Rules)\\
    AMALFI \cite{sejfia2022practical} & Package & Static (ML) \\
    MALOSS \cite{DuanAKESL21} & Package & Static \& Dynamic \\
    \textsc{Donapi} & Package & Static \& Dynamic \\
    \bottomrule
    \end{tabular}
    \end{threeparttable}
}
\vspace{-0.5em}
\end{table}

\textbf{Limitations.} 
We know that no system is entirely foolproof, and neither are our detectors. As our strategy tries to use appropriate sub-detectors for possible attack surfaces during the package installation and import stages, this leads to some inevitable limitations in \textsc{Donapi}. First, we currently only have parsers for the bash and sh commands, which are the main shell syntax used by attackers, so we need to continue to extend for other shell commands for specific scenarios (e.g., Windows shell). In addition, we set a timeout to prevent the detector from being stuck waiting for a long time, but this can lead to partial misses, such as attackers using code obfuscation and \textit{sleep} methods to escape detection, so we also need to handle such behavior during dynamic detection. Further, we found that some packages need specific conditions (e.g., environment variables \cite{Tyler2021Prevent, Jimmy2021How} and external factors \cite{Liran2021war}) in dynamic execution, and their absence will lead to a decrease in code coverage, making it impossible to accomplish dynamic monitoring accurately and generating false negatives.

Finally, distinct from the testing scenario, too many false positives can lead to significant manual review costs due to the massive volume of software packages in real-world situations, which is reflected in RQ3. Therefore, we should adjust the hierarchical classification framework according to the specific usage scenarios, e.g., if we focus on ensuring the security of the environment, we should relax the policy, i.e., more sound; if we focus on the accuracy of the report, we should tighten the policy, i.e., more complete.

\section{Related Work}
\label{sec: Related Work}
Numerous studies \cite{fang2020detecting, fass2019jstap, fang2022jstrong} for detecting malicious JavaScript code exist. However, our research explicitly targets the detection of malicious packages within the npm ecosystem rather than solely focusing on JavaScript file detection or vulnerability detection. Therefore, we will emphasize relevant work that aligns with our research direction. We have categorized these research techniques into three distinct categories.

\textbf{Machine learning.} Machine learning has emerged as a well-established approach in various research domains, including malicious package detection. Garrett et al. \cite{garrett2019detecting} proposed an anomaly detection technique for identifying and flagging anomalous update behavior since attackers tend to deviate from the regular pattern in terms of timing and frequency when using developer credentials to update packages associated with them. Liang et al. \cite{liang2023needle} are similar, except they focus on differences in API call sequences and identify anomalous packages through cluster analysis. Wyss et al. \cite{wyss2022fork} proposed a machine learning-based package difference metric in a related study designed to identify packages that share the same attack code or vulnerabilities, thus mitigating malicious code reuse. Ladisa et al. \cite{ladisa2023feasibility} proposed using language-independent features and attempted to train monolingual and cross-language models using algorithms such as XGBoost, discovering 58 previously unknown malicious packages.

Another noteworthy contribution in this field comes from Ohm et al. \cite{ohm2020supporting}. Their approach involves clustering malicious packages and utilizing code similarity to detect potentially malicious packages. However, the work closest to our goal is conducted by Sejfia et al. \cite{sejfia2022practical}. Their research uses metadata and static APIs as features, enabling classifiers for package detection. This approach offers a promising method for accurately identifying malicious packages.

\textbf{Permission system.} At the core of this approach lies program analysis and privilege control. These studies emphasize the vital role of program analysis and privilege control in identifying and mitigating potential threats associated with malicious packages. Ferreira et al. \cite{ferreira2021containing} introduced a permission system based on their investigation of package updates \cite{garrett2019detecting}. They achieved this by executing updated component packages within a sandbox environment to identify whether they contain sensitive permission calls.

Similarly, Vasilakis et al. \cite{ntousakis2021detecting} centered on evaluating third-party libraries. They utilized a combination of dynamic and static approaches to scrutinize whether the permissions associated with these packages went beyond pre-defined boundaries. Building upon this line of inquiry, they introduced a domain-specific language (DSL) specifically tailored to express Read-Write-eXecute (RWX) permissions \cite{vasilakis2021preventing}. By leveraging a detailed and granular model of RWX permissions, they effectively tackled the dynamic compromise problem, where the security of the system is compromised due to runtime modifications. In more recent work, Wyss et al. \cite{wyss2022wolf} introduce a novel permission system that evaluates malicious packages in the list and blocks any identified malicious behavior by generating a list of install-time behaviors combined with a lightweight policy language.

\textbf{Dependency analysis.} 
Dependency analysis has emerged as a prominent technique for examining code propagation paths and has found widespread application in various programming languages \cite{tang2022towards, wang2020empirical, zhan2020automated}, and JavaScript also benefits from this approach. In a study conducted by Chinthanet et al. \cite{chinthanet2021lags}, they analyzed the entire process of code updates and vulnerability fixes. The researchers identified a significant propagation delay within the npm ecosystem, potentially resulting in delayed vulnerability fixes updates. This lag raises concerns for developers and researchers alike. Similarly, Liu et al. \cite{liu2022demystifying} delved into the specific parsing rules involved in the package installation process. They proposed a knowledge graph-based dependency parsing technique that enables a more precise analysis of the dependency tree for each package. This technique facilitates better identification and tracking of vulnerability propagation. While the primary focus of these studies may not be malicious package detection, the concepts and methodologies they embody offer new perspectives and ideas for effectively detecting the propagation of malicious packages within software ecosystems.

\section{Conclusion}
\label{sec: Conclusion }
In this paper, we first build a local npm package cache containing more than 3.4 million packages for data support. Then, based on the analysis of many samples, we propose \textsc{Donapi}, an malicious npm package detector that combines dynamic and static analysis to achieve a hierarchical classification of npm malicious packages. Our detector is experimentally verified to achieve good results and has advantages over \textit{GUARDDOG}, \textit{AMAFLI} and \textit{SAP}. Ultimately, we also identified and confirmed 325 malicious packages as well as 2 API calls and 246 API call sequences never seen before, concrete proof of the value and effectiveness of our approach.

As we advance, we aim to further increase the granularity of our classification by analyzing more samples and improving the accuracy and efficiency of the classification through technological advances. Ultimately, we envision applying our techniques and methods more concretely and practically to diverse language ecosystems, not limited to the npm.

\section{Ethics and Disclosure}
\label{sec: Ethics and Disclosure }
This research uncovered real-world malicious packages. Despite the risks involved, the code is essential for a complete study presentation and has previously appeared in similar forms in the literature.

Regarding security disclosure protocol, we did not publicly disclose the malicious packages we found for security reasons. Although npm officials have unpublished most malicious packages, we hope to share them with professional security researchers via institutional email requests.

Finally, a large multinational enterprise has already deployed our detector internally for security interception, so we cannot release the code due to contract limitations and copyright issues. However, we are willing to share separately the malicious packages found in the official npm sources and their detection results for use by other researchers. We hope that our work will contribute to future research in these directions.

\section{Acknowledgements}
We would like to thank our shepherd and anonymous reviewers for their constructive comments and suggestions. This work was supported in part by National Key Research and Development Program of China (No.2021YFB3100500), Sichuan Science and Technology Program (No.2023YFG0162). 

\bibliographystyle{plain}
\bibliography{reference-simple.bib}

\section*{Appendix}
\appendix

\section{Hierarchical Classifier}
\subsection{The Importance of Sequences}
\label{Appsubsec:importance}

\begin{table}[h]
\centering
\caption{The performance on Malicious software import (M3) using sets and sequences}
\label{table:performance on sets and seqs}
\resizebox{\linewidth}{!}{
    \begin{tabular}{ccccc}
    \toprule
    
    \textbf{Category} & \textbf{APIs Representation} & \textbf{TP} & \textbf{FP} & \textbf{Instances}\\
    
    \midrule
    \hline

    \multirow{3}{*}{\makecell{Malicious \\ software import\\ (M3)}} 
    & Sets & 56 & 44 & \makecell[l]{FP: ping-me-maybe@0.0.0,\\create-sanity@3.11.5} \\
    \cline{2-5}
    ~ & Sequences & 54 & 26 & \makecell[l]{FN: noblox.js-proxies@1.0.0,\\noblox.js-proxies@1.0.3} \\ 
    \hline
    
    \bottomrule
    \end{tabular}
}
\end{table}

We obtained 100 packages from malicious datasets and real-world scenarios containing sets corresponding to our defined M3 sequences for testing. Table \ref{table:performance on sets and seqs} shows that API call sequences have fewer false positives (18) than API sets, although there are a few misses (only 2).

\subsection{Hierarchical Classification for Shell Command}
\label{Appsubsec:hierarchical}

\begin{table}[h]
\caption{YARA rules for capturing sensitive behavior of bash commands and shell scripts}
\label{table:MSCD YARA Rules}
\resizebox{\linewidth}{!}{
    \begin{threeparttable}
    \begin{tabular}{clc}
    \toprule
    \textbf{Rule} & \textbf{Description} & \textbf{Examples} \\
    \midrule
    R1 & \makecell[l]{Capturing Executable Files} & /bin/bash, /bin/sh \\
    \hline
    R2 & \makecell[l]{Capturing file manipulation commands} & scp, cat, binary, chmod\\
        \hline
    R3 & Capturing Commands for Obtaining Sensitive Information & whoami, hostname, pwd \\
        \hline
    R4 & Capturing Networking Commands & wget, curl, nc, nslookup \\    \hline
    R5 & Capturing sensitive commands & base64, b64 \\    \hline
    R6 & Capturing Sensitive Files or Folders & etc/rc, etc/passwd, /.profile\\    \hline
    R7 & Capturing commands that execute .exe file& .*.exe \\    \hline
    R8 & Capturing Suspicious Files & .sh, .exec\\
    \bottomrule
    \end{tabular}
    \end{threeparttable}
}
\end{table}

\begin{table}[h]
\caption{Mapping rules for malicious command behavior}
\label{table:MSCD rule set}
\resizebox{\linewidth}{!}{
    \begin{threeparttable}
    \begin{tabular}{ccc}
    \toprule
    \textbf{Module Name} & \textbf{Category} & \textbf{Rule Sequences} \\
    \midrule
    \multirow{4}*{\makecell{Malicious \\ shell  command \\ detector}}
      & Sensitive information theft (M1) & [R4 $\rightarrow$ R3, R4 $\rightarrow$ R6]\\
    ~ & Sensitive file operation (M2) &  [R2 $\rightarrow$ R6, R5 $\rightarrow$ R1]\\
    ~ & Malicious software import (M3) &  [R4 $\rightarrow$ R8, R7]\\
    ~ & Reverse shell (M4) &  [R4 $\rightarrow$ R1]\\
    \bottomrule
    \end{tabular}
    \end{threeparttable}
}
\end{table}
As depicted in Table \ref{table:MSCD YARA Rules}, we formulate a set of rules to capture suspicious behaviors within the commands. Leveraging these rules, we then devise a rule combination analogous to API call sequences, enabling precise categorization of both the associated malicious commands and their respective packages. This categorization is illustrated in Table \ref{table:MSCD rule set}.

\section{Obfuscation Feature Evaluation}
\label{Appsec:Obfuscation}

\begin{figure}[h]
    \centering
    \includegraphics[width=0.8\linewidth]{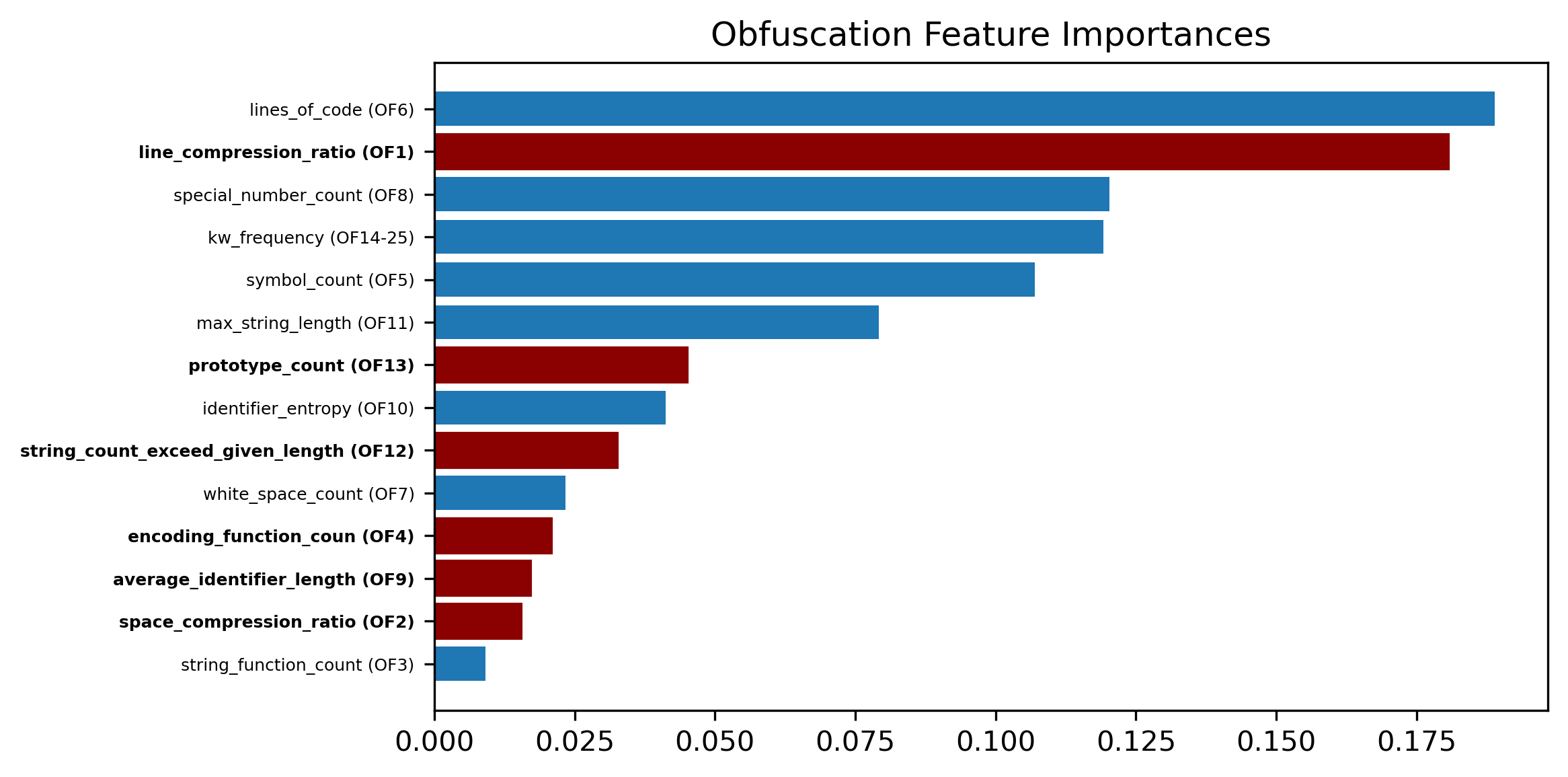}
    \caption{Obfuscation feature importance rankings. The bolded parts are new features that we propose.}
    \label{fig:Obfuscation feature evaluation}
    \vspace{-1.3em}  
\end{figure}

\section{Additional experiments on our detectors}

In addressing RQ1, we scrutinized our detectors' accuracy via a validation set, yielding a commendable performance. Nevertheless, considering that both training and validation sets originated from our accumulated malicious samples (though without overlap), these experiments inherently carry the potential for certain metric inflation. Consequently, to substantiate the stability of our detectors' accuracy in the face of sample variations, we deliberately curated distinct test sets for each subdetector and for \textsc{Donapi} independently. Our focus here is solely on the capability to detect unknown malicious or obfuscated samples, i.e., the Recall.

\subsection{Dataset}
\textbf{Local cache.} Given the inherent variability and diversity of malicious samples and to verify the validity of the local cache, we collected meta-information on 151 malicious packages disclosed by the Synk database in May and June. Subsequently, we tried and obtained all the original \texttt{.tgz} files from the local cache (not available at the official NPM), effectively proving that we can retain the repository deletion copies promptly.

\noindent
\textbf{Obfuscated code detector.} Considering that different obfuscation means may have different characteristics \cite{skolka2019anything}, we randomly selected 66 obfuscated packages from daily synchronization as obfuscated samples.

\noindent
\textbf{Static identifier \& \textsc{Donapi}.} Due to the static recognizer's lack of ability to detect commands, we excluded a subset of malicious packages (8 in total) during this testing phase. In contrast, for the \textsc{Donapi} test, we used the complete set of 151 malicious samples.

\subsection{Result}

\begin{table}[h]
\centering
\caption{Additional evaluation results of recall for sub-detectors and \textsc{Donapi}}
\label{table:Evaluation results for accuracy supply}
\resizebox{\linewidth}{!}{
    \begin{threeparttable}
    \begin{tabular}{ccccc}
    \toprule
    \textbf{Idx} & \textbf{Module Name} & \textbf{Malicious/Obfuscated} & \textbf{FN} & \textbf{Recall}\\
    \midrule
    \#1 & Obfuscated code detector & 66 & 2 & 96.97\% \\
    \#2 & Suspicious package static identifier & 143 & 8 & 94.41\%\\
    \#3 & \textbf{\textsc{Donapi} (Integral detector)} & \textbf{151} & \textbf{8} & \textbf{94.70\%} \\
    \bottomrule
    \end{tabular}
    \end{threeparttable}
}
\end{table}

As shown in Table \ref{table:Evaluation results for accuracy supply}, the detector maintains a comparable recall rate for unknown samples.

\section{API \& Behavior}


We focus on 132 APIs (see the website for their combinations with parameters). At the same time, we outline abstractions for 12 different behavior types, with each behavior type precisely defined and described. For comprehensive insights, please refer to \url{https://github.com/das-lab/Donapi}.



\end{document}